\documentclass[12pt,twoside,epsf]{article}
\usepackage{epsfig}
\usepackage{color}
\usepackage{bbold}
\usepackage{graphics}
\usepackage{booktabs}
\usepackage{amssymb}
\font\tenbf=cmbx9 %

\font\fourbf=cmbx14

\font\tenrm=cmr9

\font\tenit=cmti9 %

\font\sc=cmr12

\font\lrm=cmr9

\newcommand{\lle}{\mbox{$\langle$}}
\newcommand{\rle}{\mbox{$\rangle$}}

\newcommand{\bfsi}{\mbox{\boldmath$\sigma$}}

\newcommand{\bfep}{\mbox{\boldmath$\varepsilon$}}

\newcommand{\bfze}{\mbox{\boldmath$\zeta$}}

\newcommand{\bfchi}{\mbox{\boldmath$\chi$}}

\newcommand{\bfcK}{\mbox{\boldmath$\cal K$}}
\newcommand{\bfcL}{\mbox{\boldmath$\cal L$}}

\newcommand{\bfcP}{\mbox{\boldmath$\cal P$}}

\newcommand{\bfb}{\mbox{\boldmath$\bf b$}}

\newcommand{\bfg}{\mbox{\boldmath$\bf g$}}

\newcommand{\bfn}{\mbox{\boldmath$\bf n$}}

\newcommand{\bft}{\mbox{\boldmath$\bf t$}}
\newcommand{\bfs}{\mbox{\boldmath$\bf s$}}

\newcommand{\bfu}{\mbox{\boldmath$\bf u$}}

\newcommand{\bfx}{\mbox{\boldmath$\bf x$}}
\newcommand{\bfy}{\mbox{\boldmath$\bf y$}}
\newcommand{\bfz}{\mbox{\boldmath$\bf z$}}
\newcommand{\bfA}{\mbox{\boldmath$\bf A$}}

\newcommand{\bfG}{\mbox{\boldmath$\bf G$}}

\newcommand{\bfI}{\mbox{\boldmath$\bf I$}}

\newcommand{\bfL}{\mbox{\boldmath$\bf L$}}
\newcommand{\bfN}{\mbox{\boldmath$\bf N$}}

\newcommand{\bfX}{\mbox{\boldmath$\bf X$}}
\newcommand{\bfY}{\mbox{\boldmath$\bf Y$}}

\newcommand{\bfU}{\mbox{\boldmath$\bf U$}}

\newcommand{\bfR}{\mbox{\boldmath$\bf R$}}
\newcommand{\bfK}{\mbox{\boldmath$\bf K$}}
\newcommand{\bfM}{\mbox{\boldmath$\bf M$}}

\newcommand{\bfdelta}{\mbox{\boldmath$\delta$}}
\newcommand{\bfdel}{\mbox{\boldmath$\delta$}}
\newcommand{\bfxi}{\mbox{\boldmath$\xi$}}

\newcommand{\bfLa}{\mbox{\boldmath$\Lambda$}}
\newcommand{\bfcD}{\mbox{\boldmath$\cal D$}}

\newcommand{\bfcG}{\mbox{\boldmath$\cal G$}}

\newcommand{\bfcR}{\mbox{\boldmath$\cal R$}}
\newcommand{\bfcA}{\mbox{\boldmath$\cal A$}}
\newcommand{\bfcB}{\mbox{\boldmath$\cal B$}}

\newcommand{\cV}{\mbox{$\cal V$}}

\newcommand{\bfal}{\mbox{\boldmath$\alpha$}}
\newcommand{\bfbe}{\mbox{\boldmath$\beta$}}
\newcommand{\bfga}{\mbox{\boldmath$\gamma$}}

\newcommand{\bftau}{\mbox{\boldmath$\tau$}}

\newcommand{\bfeta}{\mbox{\boldmath$\eta$}}

\newcommand{\BB}{\begin{equation}}
\newcommand{\EE}{\end{equation}}

\newcommand{\BBEQ}{\begin{eqnarray}}
\newcommand{\EEEQ}{\end{eqnarray}}

\begin{document}


\centerline{\fourbf New RVE concept in thermoelasticity of periodic }

\centerline{\fourbf composites subjected to compact support loading}



\medskip
\vspace{12pt}
\centerline{{\bf{ Valeriy A.
Buryachenko\footnote{\tenrm Address all correspondence to V. Buryachenko:
Micromechanics and Composites, Cincinnati, OH 45202, USA; Buryach@aol.com}}}}

\vspace{4pt}
\centerline {\it Micromechanics and Composites, Cincinnati, OH 45202, USA}






\centerline{\vrule height 0.003 in width 16.0cm} \noindent
\begin{abstract}
This paper introduces an advanced Computational Analytical Micromechanics (CAM) framework for linear thermoelastic composites (CMs) with periodic microstructures. The approach is based on an exact new Additive General Integral Equation (AGIE), formulated for compactly supported loading conditions, such as body forces and localized thermal effects (e.g., laser heating). In addition, new general integral equations (GIEs) are established for arbitrary mechanical and thermal loading. A unified iterative scheme is developed for solving the static AGIEs, where the compact support of loading serves as a new fundamental training parameter. At the core of the methodology lies a generalized Representative Volume Element (RVE) concept that extends Hill’s classical definition of the RVE. Unlike conventional RVEs, this generalized RVE is not fixed geometrically but emerges naturally from the characteristic scale of localized loading, thereby reducing the analysis of an infinite periodic medium to a finite, data-driven domain. This formulation automatically filters out nonrepresentative subsets of effective parameters while eliminating boundary effects, edge artifacts, and finite-size sample dependencies. Furthermore, the AGIE-based CAM framework integrates seamlessly with machine learning (ML) and neural network (NN) architectures, supporting the development of accurate, physics-informed surrogate nonlocal operators.
\end{abstract}

\centerline{\vrule height 0.003 in width 16.0cm} \noindent

\section{Introduction }

Analytical micromechanics of random-structured composites has long relied on a few core ideas—most notably the Effective Field Hypothesis (EFH), the General Integral Equation (GIE), and the Representative Volume Element (RVE). These principles have shaped many classical models, but also impose significant constraints. In this work, we show that such assumptions, though historically central, can be relaxed or even discarded, enabling broader and more versatile approaches to micromechanical modeling.
The effective field hypothesis (EFH), introduced by pioneers such as Poisson, Faraday, Mossotti, Clausius, Lorenz, and Maxwell (see \cite{Buryachenko`2007, Buryachenko`2022}), assumes that inclusions at $\bfx \in v_i$ experience a uniform local field distinct from the remote macroscopic one. Denoted as {\bf H1a}, EFH has been the cornerstone of analytical micromechanics {\color{black}(called the first background of micromechanics}) for over 150 years, both as a conceptual framework and as an approximation for solving the general integral equation (GIE). The GIE—tracing back to Rayleigh (1892) \cite{Rayleigh`1892}—links random fields at a point with those in its surroundings. While classical GIEs implicitly rely on EFH, new operator-based formulations \cite{Buryachenko`2022} define computational analytical micromechanics (CAM, {\color{black} called also the second background of micromechanics}), independent of Green’s functions or constitutive models. CAM improves local field accuracy, even correcting sign errors inside inclusions. Traditional methods such as the Effective Field Method (EFM) and Mori–Tanaka Method (MTM) can be interpreted as particular GIE solution strategies.

The Representative Volume Element (RVE) is essential for predicting the effective behavior of heterogeneous materials. A properly sized RVE captures microstructural heterogeneity while minimizing size and boundary artifacts. Following Hill \cite{Hill`1963}, the RVE assumes macroscopically homogeneous boundary conditions, allowing a well-defined effective moduli tensor.RVE responses are typically evaluated via Direct Numerical Simulations (DNS) of Microstructural Volume Elements (MVEs) derived from synthetic models or imaging techniques such as micro-CT \cite{Echlin`et`2014, Konig`et`1991, Ohser`M`2000}. Selecting an appropriate RVE requires balancing statistical representativeness and scale separation, generally expressed as $a \ll \Lambda \ll L$, where $a$ is the microstructural size, $\Lambda$ is the characteristic field variation, and $L$ is the macroscopic scale. Under such conditions, classical homogenization reliably predicts effective properties. To validate an RVE, its response should be independent of boundary conditions, typically verified through convergence studies identifying the smallest domain yielding stable properties. The related concept of Statistically Equivalent RVE (SERVE) uses image-based, data-driven approaches to replicate microstructures for simulation. Methods for determining RVE and SERVE sizes are reviewed in \cite{Bargmann`et`2018, Francqueville`et`2019, Harper`et`2012, Kanit`et`2003, Matous`et`2017, Moumen`et`2021, Ostoja`et`2016, Sab`N`2005}.

Periodic composites, owing to their regular microstructure, are naturally suited for multiscale homogenization \cite{Fish`2014, Ghosh`2011, Zohdi`W`2008}. In asymptotic homogenization, originally developed by Babuška and formalized in \cite{Bakhvalov`P`1984, Fish`2014}, the response of a periodic medium is obtained by separating scales, assuming the unit cell is much smaller than the overall structure, leading to homogenized coefficients via unit-cell problems. Computational homogenization, in contrast, solves microstructural field equations numerically, enabling modeling of nonlocal \cite{Buryachenko`2025} and inelastic behavior \cite{Kouznetsova`et`2001, Matous`et`2017, Terada`K`2001}. A key framework is FE$^2$, where macroscopic finite elements are coupled to microscale RVEs at Gauss points \cite{Geers`et`2010, Kanout`et`2009, Raju`et`2021}, each solving boundary value problems consistent with the macroscopic deformation, usually under periodic BCs. While rigorous, FE$^2$ is computationally expensive due to nested discretizations, often requiring high-performance computing or model reduction for large-scale simulations
When the scale separation assumption fails, the hypothesis of statistically homogeneous fields is no longer valid. Consequently, stress and strain averages become coupled in a nonlocal manner through a tensorial kernel, requiring the use of effective elastic operators written in integral form. Such operators include, following the classification in \cite{Maugin`2017}, strongly nonlocal theories (e.g., strain- and displacement-based approaches such as peridynamics by Silling \cite{Silling`2000}), as well as weakly nonlocal models such as strain-gradient or stress-gradient formulations. Within this framework, micromechanics plays a crucial role in linking scales that are governed by nonlocal constitutive laws. Instead of the classical effective moduli introduced by Hill \cite{Hill`1963}, one must employ effective nonlocal operators, expressed either in integral or differential form. This shift leads to a generalized concept of the representative volume element (RVE), applicable to both random microstructures (\cite{{Drugan`2000},{Drugan`2003},{Drugan`W`1996}}) and periodic composites (\cite{Ameen`et`2018}, \cite{Kouznetsova`et`2004a}, \cite{Kouznetsova`et`2004b}, \cite{Smyshlyaev`C`2000}). The generalized RVE is essential for capturing nonlocal effects arising from heterogeneous fields, intrinsic material nonlocality, and long-range inclusion interactions. In conventional treatments, both classical and generalized RVEs depend on prescribed boundary conditions and a chosen form of the nonlocal operator, which can restrict generality. These limitations may be overcome by considering body forces or temperature variations of compact support (as in laser heating of composites), which define a new class of loading scenarios. Such problems are closely related to those in functionally graded materials (FGMs) and are effectively addressed by the GIE-CAM framework \cite{Buryachenko`2022}. In general, this requires solving the GIE in the full space. However, the problem simplifies considerably with the new Additive General Integral Equation (AGIE), tailored for localized loading. In AGIE, the free term represents the deformation of an infinite homogeneous matrix and is given in explicit form, while the renormalization term in the original GIE is omitted. As a result, the AGIE solution appears as a streamlined variant of the classical GIE formulation. A key advantage of AGIE is its redefinition of the RVE: rather than an infinite domain (as in Hill’s classical concept \cite{Hill`1963}), the RVE is determined solely by the compact support of loading. It is thus identified as the region outside which strains and stresses vanish. This provides a fundamentally new and physically meaningful perspective on the RVE in the presence of nonlocal behavior and localized excitations.

The development of effective nonlocal operator theory has been significantly advanced by the integration of machine learning (ML) and neural networks (NN), offering enhanced flexibility and predictive power. Early work by Silling \cite{Silling`2020} and You {\it et al.} \cite{You`et`2020, You`et`2024} illustrated how direct numerical simulation (DNS) data can be leveraged to build surrogate integral operators for complex materials. More recently, neural operators have been introduced to learn mappings between function spaces \cite{Li`et`2003, Lanthaler`et`2024}, with architectures such as DeepONet, PCA-Net, Graph Neural Operators, FNO, and LNO designed for different operator-learning tasks (see reviews in \cite{Gosmani`et`2022, Hu`et`2024, Kumara`Y`2023, Lanthaler`et`2024}).
In nonlocal mechanics, the Peridynamic Neural Operator (PNO) \cite{Jafarzadeh`et`2024} and its heterogeneous variant HeteroPNO \cite{Jafarzadeh`et`2024b} provide physics-aware modeling of peridynamic interactions. Physics-Informed Neural Networks (PINNs) further incorporate governing equations as soft constraints \cite{Raissi`et`2019, Karniadakis`et`2021, Hu`et`2024}, ensuring physically consistent predictions. Combining neural operators with PINNs \cite{Faroughi`et`2024, Gosmani`et`2022, Wang`Y`2024} enables accurate modeling of complex, nonlinear, heterogeneous, and nonlocal material behavior with strong generalization. 
While ML\&NN methods have greatly expanded the scope of material modeling, they often neglect key micromechanical principles—such as scale separation, boundary influences, and the rigorous definition of RVE—that are essential for predictive accuracy in both linear and nonlinear regimes. To overcome these shortcomings, a new methodology is introduced that leverages compressed datasets specifically structured for complex microstructures within a redefined RVE framework. Unlike conventional approaches that depend on phase-specific constitutive laws or explicit analytical surrogate operators, this framework characterizes microstructures through field concentration factors within each constituent phase, ensuring a physically consistent representation. The resulting datasets, derived from this micromechanically grounded RVE, are broadly compatible with diverse ML and NN architectures for constructing nonlocal surrogate operators. By mitigating challenges related to size dependence, boundary effects, and artificial edge artifacts, this RVE-based strategy significantly improves the robustness and accuracy of surrogate predictions across different material systems and loading conditions.

The paper is structured as follows. Section 2 outlines the governing equations and notation of elasticity, including the treatment of body forces and thermal effects with compact support, together with two types of periodic boundary conditions (PBC). Section 3 examines the response of an infinite homogeneous matrix under localized loading, presenting decompositions of material parameters and field variables, as well as the General Integral Equations (GIEs) and their additive form (AGIEs), which account for both average fields and matrix-induced fields. Section 4 discusses classical and newly proposed RVE concepts. Section 5 develops the framework of compressed datasets and a redefined RVE for periodic composites under compact support body forces and thermal changes. It further introduces the construction of effective elastic moduli and a class of surrogate nonlocal operators, trained using localized loading conditions as inputs.


\section{Preliminary}
\setcounter{equation}{0}
\renewcommand{\theequation}{2.\arabic{equation}}

\subsection{Basic equations}

We study a linear thermoelastic body occupying a bounded, simply connected domain $w\subset R^d$ with smooth boundary $\Gamma_0$ and with an indicator function $W$ and space dimensionality
$d$ ($d=2$ and $d=3$ for 2-$D$ and 3-$D$ problems, respectively). The medium consists of
a homogeneous matrix phase $v^{(0)}$, and
a periodic array of heterogeneities $X=(v_i)$ {\color{black}(that share identical mechanical and geometric properties)}, where inclusion $v_i$ is centered at $\bfx_i\in \bfLa$ of the grid $\bfLa$, carries indicator $V_i$, and is bounded by a smooth closed surface $\Gamma_i$ $(i=1,2,\ldots)$; {\color{black} i.e. two-phase CMs are considered}.
We consider the local basic equations of thermoelasticity of composites
\BBEQ
\label{2.1}
\nabla\cdot\bfsi(\bfx)&=&-\bfb(\bfx), \\ 
\label{2.2}
\bfsi(\bfx)&=&\bfL(\bfx)\bfep(\bfx)+\bfal(\bfx), \ \ {\rm or}\ \ \
\bfep(\bfx)=\bfM(\bfx)\bfsi(\bfx)+\bfbe(\bfx), \\ 
\label{2.3}
\bfep(\bfx)&=&\nabla^s\bfu, \ \
\nabla\times\bfep(\bfx)\times\nabla={\bf 0}, 
\EEEQ
where $\otimes$ and $\times$ denote the tensor and vector products, respectively;
$\nabla^s$ stands for the symmetric gradient operator, defined as
$\nabla^s\bfu:=[\nabla {\otimes}{\bf u}+(\nabla{\otimes}{\bf u})^{\top}]/2$, with $(.)^{\top}$ indicating matrix transposition.
The vector $\bfb$ represents the body force.
The tensors ${\bf L}(\bfx)$ and ${\bf M}(\bfx) \equiv {\bf L}(\bfx)^{-1}$ correspond to the known phase stiffness and compliance tensors.
The second-order tensors $\bfbe(\bfx)$ and $\bfal(\bfx) = {\color{black}\bfL(\bfx)\bfbe(\bfx)}$ describe the local eigenstrains and eigenstresses, respectively.
In the special case of isotropic phases, the stiffness tensor $\bfL(\bfx)$ can be expressed in terms of the local bulk modulus $k(\bfx)$ and shear modulus $\mu(\bfx)$ as follows:
\BB
\label{2.4}
\bfL(\bfx)=(dk,2\mu)\equiv dk(\bfx)\bfN_1+2\mu(\bfx)\bfN_2, \ \ \bfbe(\bfx)=\beta^{t}\theta\bfdel,
\EE
${\bf N}_1=\bfdelta\otimes\bfdelta/d, \ {\bf N}_2={\bf I-N}_1$ $(d=2\ {\rm or}\ 3$) ,
where $\bfdelta$ and $\bfI$ denote the unit second- and fourth-order tensors, respectively.
The quantity $\theta = T - T_0$ represents the temperature change relative to a reference temperature $T_0$, and $\beta^{t}$ is the thermal expansion coefficient.
For any material tensor $\bfg$ (e.g., $\bfL, \bfM,\bfbe,\bfal)$ , we use the notation
$\bfg_1(\bfx)\equiv \bfg(\bfx)-\bfg^{(0)}=\bfg^{(m)}_1(\bfx)$ $(\bfx\in v^{(m)},\ m=0,1$).
Here, the superscript $^{(m)}$ indicates the material component, while the subscript $i$ refers to individual heterogeneities.
We define $v^{(0)}=w\backslash v$, $ v\equiv \cup v_i,
\ {\color{black}V(\bfx)=\sum V_i(\bfx)}$, where $V_i(\bfx)$ are the indicator functions of $v_i$: equal to 1 for $\bfx \in v_i$ and 0 otherwise $(i=1,2,\ldots)$.
By substituting Eqs. (\ref{2.2}) and (\ref{2.3}) into Eq. (\ref{2.1}), the equilibrium equation (\ref{2.1}) can be rewritten in the following form:
\BB
\label{2.5}
{\bfcL}(\bfu,\bfal)(\bfx)+\bfb(\bfx)={\bf 0},\ \ \ {\bfcL}(\bfu,\bfal)(\bfx):={\color{black}\nabla\cdot}[\bfL\nabla\bfu(\bfx)+\bfal(\bfx)],
\EE
where ${\bfcL}(\bfu,\bfal)(\bfx)$ denotes a second-order elliptic differential operator.
In what follows, to streamline the calculations, we introduce the following substitutions:
\BB
(\bfep,\bfsi) \leftrightarrow \bfxi,\ \ (\bftau,\bfeta) \leftrightarrow \bfze,\ \ (\bfal,\bfbe) \leftrightarrow \bfga
\label{2.6}
\EE
{\color{black} indicating that, for each first $\bfxi=\bfep,\ \bfze=\bftau, \ \bfga=\bfal$ 
and second $\bfxi=\bfsi,\ \bfze=\bfeta, \ \bfga=\bfbe$ triplet, the subsequent equations in the variables $\bfxi\ \bfze,$ and $\bfga$ reduce to the corresponding relations for the strain $\bfep$ 
(first triplet) and $\bfsi$ (second triplet).}
The stress and strain polarization tensors (introduced merely as a notational convenience)
\BB
\label{2.7}
\bftau(\bfx)=\bfL_1(\bfx)\bfep(\bfx)+\bfal_1(\bfx), \ \ \bfeta(\bfx)=\bfM_1(\bfx)\bfsi(\bfx)+\bfbe_1(\bfx),
\EE
($\bfeta(\bfx)=-\bfM^{(0)}\bftau(\bfx)$),
respectively, vanishing in the matrix $\bftau (\bfy) \equiv \bfeta(\bfy)\equiv{\bf 0}$ ($\bfy\in v^{(0)}$).

The body force density $\bfb(\bfx)$ is assumed to be of compact support (referred to as a body force with compact support, BFCS), self-equilibrated, and vanishing outside a prescribed loading region $\bfcB^b := b(\mathbf{0}, B^b)$:
\BB
\label{2.8}
\int \bfb(\bfx)d\bfx={\bf 0}, \ \ \ \bfb(\bfy)\equiv {\bf 0}\ \ {\rm for}\ \ \bfy\not\in b({\bf 0}, B^b):=\{\bfy| |\bfy|\leq B^b\},
\EE
where $b(\mathbf{0}, B^b)$ denotes the ball of radius $B^b$ centered at the origin $\bfx = \mathbf{0}$.
Similarly, the temperature {\color{black} change} $\theta$ with compact support (TCCS) in the domain $\bfcB^{\theta} := b(\mathbf{0}, B^{\theta})$ is defined by
\BB
\label{2.9}
\theta(\bfx)\equiv {\bf 0}\ \ {\rm for}\ \ \bfy\not\in b({\bf 0}, B^{\theta}):=\{\bfy| |\bfy|\leq B^{\theta}\},
\EE
i.e. $\bfal(\bfx),\bfbe(\bfx)$ have the same compact support $\bfcB^{\theta}$.

\subsection{Periodic structure CMs and some averages}

For notational clarity, we examine a two-dimensional periodic composite body, subject to a body force density $\bfb(\bfx)$ (\ref{2.5})
(compactly supported, self-equilibrated, and periodic within a body-force unit cell $\Omega^b_{00}\supset \bfcB^b$, which covers the region $\bfcB^b$. For simplicity, we assume that $\bfcB^b=\bfcB^{\theta}$ and $\Omega^b_{00}$ are identical for $\bfb(\bfx)$ and $\theta(\bfx)$. 
The domain $w$ is modeled as an infinite tessellation of square unit cells,
$w=\cup \Omega_{ij}^b$ ($i,j=0,\pm 1, \pm 2,\ldots$), 
where each cell $\Omega_{ij}^b$ corresponds to the translation by the periodic lattice $\bfLa^b=\{\bfx^{b\Lambda}\}$.
A representative {\color{black} body force unit cell (BFUC) } $\Omega^b_{00}$ is defined with corner points $\bfx^{bc}_{kl}$ ($k,l = \pm1$), and a piecewise boundary $\Gamma^{b0} = \cup \Gamma^{b0}_{ij}$. Each segment $\Gamma^{b0}_{ij}$ serves as the interface between 
{\color{black} $\Omega^b_{00}$ and its neighboring cell $\Omega^b_{ij}$. Here, the indeces $(i,j)$} 
satisfy $i = 0, \pm1$ and $j = \pm(1 - |i|)$ (see Fig. 1 reproduced from \cite{Buryachenko`2022}). The entire grid of such unit cells respects periodic repetition, so that $\Omega^b_{00}$ deforms and responds identically to all neighboring cells, ensuring overall periodicity of the structure—much like the conceptual tiling seen in simulations using periodic loading (mechanical and thermal)
\BB
\label{2.10}
\bfb(\bfx-\bfchi)=\bfb(\bfx),\ \ \theta(\bfx-\bfchi)=\theta(\bfx), \ \ \bfchi\in {\bfLa}^b.
\EE

\vspace{-7.mm} \noindent \hspace{30mm}
\parbox{8.8cm}{
\centering \epsfig{figure=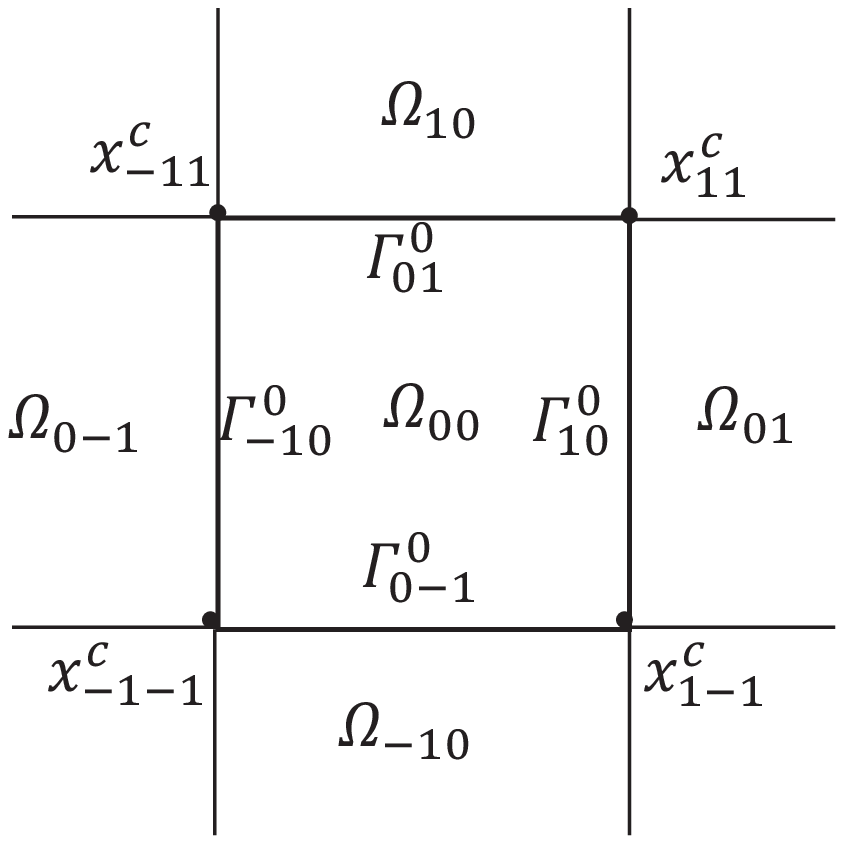, width=7.8cm}\\ \vspace{-50.mm}
\vspace{132.mm}
\vspace{-125.mm} \tenrm \baselineskip=8pt
{{ Fig. 1:} A unit cell $\Omega_{00}$ with the boundary $\Gamma^0_{ij}$ and the
surrounding UCs $\Omega_{ij}$}}
\vspace{0.mm}

For the loading (\ref{2.10}), Eq. (\ref{2.5}) are supplemented by the periodic boundary conditions (PBC) at the body force unit cell (BFUC) 
$\Omega_{00}^b$: 
\BB
\label{2.11}
\bfep(\bfx) \#, \ \ \ \bfsi(\bfx)\cdot \bfn(\bfx)=-\#, \ \ \bfx\in \partial \Omega^b_{00}
\EE
where the strain field $\bfep(\bfx)$ is prescribed as periodic (denoted $\bfep \#$), while the associated traction vector $\bfsi(\bfx)\cdot\bfn(\bfx)$ is enforced to be anti-periodic (denoted $\bfsi\cdot\bfn=-\#$). Such boundary requirements guarantee both compatibility and equilibrium across the interfaces of neighboring periodic cells. 
The periodic boundary conditions (PBC, \ref{2.11}) in the case of 
\BB
\label{2.12}
{\color{black}\bfb(\bfx)\equiv{\bf 0} \ \ {\rm and}\ \ \theta(\bfx)\equiv 0}
\EE
reduce to the well-known homogeneous remote loading conditions, which are equivalently referred to as kinematic uniform boundary conditions (KUBC) and static uniform boundary conditions (SUBC), imposed through appropriate constant symmetric tensors $\bfep^{w_{\Gamma}}$ or $\bfsi^{w_{\Gamma}}$: 
\BBEQ
\bfu(\bfy)&=& \bfep^{w_{\Gamma}}\bfy, \ \forall\bfy\in \Gamma_{0u}={\Gamma}_0,\nonumber\\
\label{2.13}
\bft(\bfy)&=&\bfsi^{w_{\Gamma}}\bfn(\bfy), \ \forall\bfy\in \Gamma_{0\sigma}={\Gamma}_0,
\EEEQ
which correspond to the analysis of the governing equations in terms of either strain or stress, respectively. The two formulations are formally analogous.

It is assumed that the composite medium
$w=\cup \Omega_{ij}$ ($i,j=0,\pm1,\pm2,\ldots$) possesses a periodic microstructure, where each unit cell (UC) is represented by the domain $\Omega_{ij}$, with its lattice of centers given by $\bfLa={\bfx_{ij}}$. The compact support loading $\bfb(\bfx)$ and $\theta(\bfx)$ are introduced over an enlarged region $\Omega^b_{00}$, which is chosen to contain the reference unit cell $\Omega_{00}\subset\Omega^b_{00}$
\BB
\label{2.14}
{\color{black}\bfL(\bfx-\bfchi)=\bfL(\bfx),\ \ \ \bfchi\in {\bfLa},}
\EE
i.e. two periodic latices $\bfLa^b$ and $\bfLa$ are considered. 
For the loading fields $\bfb(\bfx)$ (\ref{2.8}) and $\theta(\bfx)$ (\ref{2.9}), the periodic boundary conditions (PBC) (\ref{2.11}) remain applicable on the enlarged domain $\Omega^b_{00}$. In contrast, the classical periodic conditions on the smaller reference unit cell (UC) $\Omega_{00}$ at the material scale,
\BB
\label{2.15}
\bfep(\bfx) \#, \ \ \ \bfsi(\bfx)\cdot \bfn(\bfx)=-\#, \ \ \bfx\in \partial \Omega_{00}
\EE
are in general {\it not satisfied}, owing to the extended support of the loading (\ref{2.11}), which exceeds the boundaries of $\Omega_{00}$. Consequently, equilibrium across neighboring cells $\Omega_{ij}$ is no longer guaranteed, and the formulation must instead be posed on the enlarged subdomain $\Omega^b_{ij}$ to preserve consistency.
{\color{black} The rationale for introducing two distinct periodic lattices—one associated with the loading and the other with the material—is clarified in Comment 4.2; a graphical illustration for the 2D case is provided in Fig. 4a.}

We distinguish between two fundamentally different problem settings.
The first problem $\bfcR^{\rm iso}$ {\color{black} (isolated, corresponding to isolated domain
$\Omega_{00}^b$)} 
concerns the governing equation (\ref{2.5}) defined on a single isolated domain $\Omega_{00}^b$, where the loading is provided in the form of a compact support loading BFCS (\ref{2.8}) and TCCS (\ref{2.9}). In this case, the external boundary of $\Omega_{00}^b$ is assumed to be traction-free, and therefore the problem represents the local response of a finite body subjected to a localized loading distribution.

The second problem $\bfcR^{\rm per}$ {\color{black} (periodic, corresponding to the periodic lattice $\bfLa^b$)} addresses the same governing equation (\ref{2.5}), but now posed in the context of an infinite periodic composite medium. Here, the applied loading is given by the periodic loading (\ref{2.10}), and the corresponding boundary conditions are taken as the standard periodic boundary conditions (PBCs) (\ref{2.11}). 
In Section 4, we return to a comparative analysis of these two formulations and explicitly identify the assumptions under which their solutions become equivalent. More precisely, we will show that the displacement and strain fields obtained within the isolated finite domain $\Omega_{00}^b$ (problem $\bfcR^{\rm iso}$) under compactly supported loading coincide with those found in the corresponding representative unit domain $\Omega_{00}^b$ embedded in the infinite periodic material {\color{black} (problem 
$\bfcR^{\rm per}$)}. This equivalence highlights the precise conditions necessary for transferring results from a localized finite-domain problem to the framework of a fully periodic homogenization setting.

We focus on a widely studied class of composites—matrix composites—consisting of a continuous matrix phase reinforced by isolated inhomogeneities of various shapes.
Three characteristic material length scales are introduced (see, e.g., \cite{{Torquato`2002}, {Malyarenko`O`2019}}):
the macroscopic scale $L$, associated with the size of the domain $w$, the microscopic scale $a$, defined by the heterogeneities $v_i$.
In addition, the applied field is assumed to vary over a characteristic length scale $\Lambda$.
The ranges of interest for both material and field scales are given by
\BBEQ
\label{2.16}
\!\!\!\!\!\!\!\!\!\!\!\!\!\!\!L\geq\Lambda\geq a^{\rm int}\geq a\ \ \ \!\!\!\!\!\!\!\! &{\rm or}&\!\!\!\!\!\!\!\!\ \ \ L\gg \Lambda\gg a^{\rm int}\geq a,\\
\label{2.17}
\!\!\!\!\!\!\!\!\!\!\!\!\!\!\!\!\!\!\!\!L\geq \Lambda\geq|\Omega_{00}| \ \ \ \!\!\!\!\!\!\!\!&{\rm or}&\!\!\!\!\!\!\!\! \ \ L\gg \Lambda\gg|\Omega_{00}|,
\EEEQ
where the inequalities (\ref{2.16}$_2$) and (\ref{2.17}$_2$) correspond to the scale separation hypotheses. The relations in (\ref{2.16}) describe the case of random-structure composites, where $a^{\rm int}$ denotes the scale of long-range interactions between inclusions (e.g., $a^{\rm int} = 6a$). The relations in (\ref{2.17}) characterize periodic-structure composites, where $\Omega_{00}$ is the unit cell. For brevity, we write $|\Omega_{00}|$ instead of the linear measure $|\Omega_{00}|^{1/d}$.

For periodic structure CMs, the notation
${\bf f}^{\small \Omega}(\bfx)$
will be used for the average of the function ${\bf f}$
over the cell $\bfx\in\Omega_{ij}$ with the center
$\bfx^{\Omega}_{ij}\in {\color{black}\Omega_{ij}}$:
\BB
{\bf f}^{\Omega}(\bfx)={\bf f}^{\Omega}(\bfx_i^{\Omega})\equiv
n(\bfx)\int_{\Omega_{ij}}{\bf f}(\bfy)~d\bfy,\quad \bfx\in \Omega_{ij},
\label{2.18}
\EE
$n(\bfx)\equiv 1/\overline \Omega_{ij}$
is the number density of inclusions in the cell $\Omega_{ij}$.

Let $\cV_{\bf x}$ denote a “moving averaging” cell (or moving window \cite{{Buryachenko`2022},{GrahamBrady`et`2003}}) centered at $\bfx$, with characteristic size $a_{\cV}=({\overline{\cV}})^{1/d}$.
For definiteness, consider a random vector $\bfchi$ uniformly distributed over $\cV_{\bf x}$, such that its value at a point $\bfz \in \cV_{\bf x}$ is
$\varphi_{\small \bfchi}(\bfz)=1/\overline {\cV}_{\bf x}$ and $\varphi_{\small \bfchi }(\bfz)\equiv 0$ otherwise. 
Then, the average of a function ${\bf f}$ with respect to translations of $\bfchi$ is defined as
\BB
\langle {\bf f} \rangle _{\bf x}(\bfx-\bfy)={1\over\overline
{\cV}_{\bf x}}\int_{\cV_{\rm \bf X}}{\bf f}(\bfz -\bfy)~d\bfz,
\quad \bfx\in {\Omega} _i.
\label{2.19}
\EE
The ``moving averaging" cell $\cV_{\bf x}$ can, in general, be obtained by translating a reference cell $\Omega_{ij}$, and its size and shape may vary from point to point.
For clarity of exposition, we assume that $\cV_{\bf x}$ arises from $\Omega_{ij}$ through a translation by the vector $\bfx-\bfx^{\Omega}_{ij}$.

{\color{black}
Interestingly, most subsequent results for periodic media were first obtained for random composites described by probability density functions, since periodic structures are special cases of random composites. 
Indeed, the random structure CMs 
(see for detals and references \cite{Buryachenko`2022}) 
are described by a conditional
probability density $\varphi (v_i,{\bf x}_i \vert v_1,{\bf x}_1$
for finding a heterogeneity of type $i$ with the center $\bfx_i$ in the domain $v_i$, with the fixed heterogeneities $v_1$ centered at ${\bf x}_1$.
The notation $\varphi (v_i , {\bf x}_i\vert ;v_1,{\bf x}_1)$ denotes the case ${\bf x}_i\neq
{\bf x}_1$. We have $\varphi (v_i, {\bf x}_i\vert ;v_1,{\bf x}_1)\to \varphi(v_i, {\bf x}_i)$
as $\vert {\bf x}_i-{\bf x}_1\vert\to \infty$ (since no long-range order is assumed), and 
$\varphi (v_i,{\bf x}_i)$ is a number density, $n^{(k1}=n^{(1)}({\bf x})$ of component $v^{(1)}$ at the point ${\bf x}$. Then any deterministic (e.g., periodic) field of inclusions $v^{(1)}$ with the centers $\bfx_{\alpha}\in\bfLa$ can be presented by
the probability density $\varphi (v_i,{\bf x}_i )$ and conditional probability density
$\varphi (v_i,{\bf x}_i \vert; v_j,{\bf x}_j)$ expressed through the $\delta$ functions ($\bfx_{\bf \alpha}\in \bfLa$)}
\BBEQ
{\color{black}\varphi (v_i,{\bf x}_i )}&=&{\color{black}\sum_{\bf \alpha}\delta (\bfx_i-\bfx_{\bf \alpha})}, \nonumber\\
\label{2.20}
{\color{black}\varphi (v_i,{\bf x}_i\vert; v_j,\bfx_j)}&=&{\color{black}\sum_{\bf \alpha}\delta (\bfx_i-\bfx_{\bf \alpha})-\delta(\bfx_i-\bfx_j).}
\EEEQ

\section{Additive general integral equations (AGIE)}
\setcounter{equation}{0}
\renewcommand{\theequation}{3.\arabic{equation}}

\subsection{Infinite homogeneous matrix subjected to compact support loading}

We introduce a linear-elastic reference medium, described by a homogeneous stiffness tensor $\bfL^{(0)}$. The governing equilibrium problem is formulated for an infinite homogeneous body occupying the whole space $\mathbb{R}^d$ ($d=1,2,3$), subjected to the body force density $\bfb(\bfx)$ specified in Eq. (\ref{2.8}):
\BB
\label{3.1}
{\bfcL}^{(0)}(\bfu^{b(0)})(\bfx)+\bfb(\bfx)={\bf 0},
\EE
where ${\bfcL}^{(0)}$ represents the elliptic differential operator determined by the constant stiffness tensor $\bfL^{(0)}$ (\ref{2.5}).
The action of this force field generates a displacement response throughout the medium, denoted by:
\BB
\label{3.2}
\bfu^{b(0)}(\bfx)\equiv - ({\bfcL}^{(0)})^{-1} \bfb.
\EE
Equivalently, the displacement field can be expressed in terms of the Green operator $\bfG^{(0)}(\bfx)$ associated with the Navier equation (\ref{2.5}) for the homogeneous reference stiffness tensor $\bfL^{(0)}$:
\BB
\label{3.3}
\bfu^{b(0)}(\bfx)= \int \bfG^{(0)}(\bfx-\bfy)\bfb(\bfy)~d\bfy.
\EE
In this representation, the Green operator $\bfG^{(0)}(\bfx)$ plays the role of the inverse of the reference differential operator, and may be interpreted physically as the fundamental solution describing the displacement response of an infinite homogeneous medium to a point (localized) body force.
Once the displacement field is determined through Eq. (\ref{3.3}), the corresponding strain and stress fields follow directly by applying the standard strain–displacement and stress–strain relations to $\bfu^{b(0)}(\bfx)$.
\BBEQ
\label{3.4}
\bfep^{b(0)}(\bfx)&=& \int \nabla^s\bfG^{(0)}(\bfx-\bfy)\bfu^{(b)}(\bfy)~d\bfy,\\
\label{3.5}
\bfsi^{b(0)}(\bfx)&=& \bfL^{(0)} \int \nabla^s\bfG^{(0)}(\bfx-\bfy)\bfu^{(b)}(\bfy)~d\bfy.
\EEEQ
We next turn to the case of an infinite homogeneous medium characterized by the stiffness tensor $\bfL^{(0)}$ and eigenstrain distribution $\bfbe^{(0)}(\bfx)$, subjected to zero body force, $\bfb(\bfx)\equiv {\bf 0}$, but influenced by the temperature variation $\theta$ given in (\ref{2.9}). In this setting, the field variables are naturally grouped as $(\bfep,\bfsi)\leftrightarrow\bfxi$ and $(\bfal_1,\bfbe_1)\leftrightarrow\bfga_1$, leading to the representation
\BB
\label{3.6}
\bfxi(\bfx)=\int\bfU^{(0)\gamma}(\bfx-\bfy)\bfga_1^{(0)}(\bfy)~d\bfy,
\EE
where $\bfU^{(0)\gamma}$ denotes the pair of strain–stress Green tensors, $\bfU^{(0)}$ and ${\bf\Gamma}^{(0)}$.
The strain Green tensor $\bfU^{(0)}$ is obtained as the second gradient of the displacement Green tensor $\bfG^{(0)}$, i.e.,
$U^{(0)}_{ijkl}(\bfx)
=\nabla_j\nabla_ lG^{(0)}_{(ij)(kl)}$, 
and exhibits the decay behavior $O\big(\int |\bfx|^{1-d}d|\bfx|\big)$ as $|\bfx|\to\infty$, ensuring that $\bfU^{(0)}(\bfx)\to{\bf 0}$ at infinity. The symmetrization over the lower indices (indicated by parentheses) guarantees compliance with the intrinsic symmetry conditions of elasticity.
The associated stress Green tensor is given by
${\bf \Gamma}^{(0)}=-\bfL^{(0)}(\bfI\delta(\bfx)+\bfU^{(0)}(\bfx)\bfL^{(0)})$, 
which links the stress response to localized eigenstrain sources within the infinite reference medium.

\subsection {Material and field decompositions}

The material properties are decomposed as
\BBEQ
\label{3.7}
\bfL(\bfx)=\bfL^{(0)}+\bfL_1(\bfx) , \ \ \bfga(\bfx)=\bfga^{(0)}(\bfx)+\bfga_1(\bfx),
\EEEQ
which may be interpreted as an extraction of the reference matrix characteristics, such that the corresponding property jumps vanish identically within the matrix phase $v^{(0)}$. Due to the compact support of the temperature variation $\theta(\bfx)$ (\ref{2.9}), the function $\bfga^{(0)}(\bfx)$ is generally nonuniform and hence $\bfga^{(0)}(\bfx)\not\equiv\mathrm{const}$.
An analogous strategy is employed for the decomposition of the total field quantities themselves, where the extraction procedure is carried out with respect to the mechanical response of the reference matrix under applied loading
\BBEQ
\label{3.8}
\!\!\!\!\!\!\!\!\!\!\!\!\!\!\!\bfep(\bfx)\!&=&\! \bfep^I(\bfx)\!+\!\bfep^{II}(\bfx), \nonumber\\
\bfep^I(\bfx)&=&\bfep^I_0(\bfx)\!+\!\bfep^I_1(\bfx), \
\bfep^{II}(\bfx)= \bfep^{II}_0(\bfx)\!+\!\bfep^{II}_1(\bfx),\\
\label{3.9}
\!\!\!\!\!\!\!\!\!\!\!\!\!\!\!\bfu(\bfx)\!&=&\! \bfu^I(\bfx)\!+\!\bfu^{II}(\bfx), \nonumber\\
\bfu^I(\bfx)&=&\bfu^I_0(\bfx)\!+{\color{black}\!\bfu^I_1(\bfx)},
\bfu^{II}(\bfx)= \bfu^{II}_0(\bfx)\!+\!\bfu^{II}_1(\bfx),\\
\label{3.10}
\!\!\!\!\!\!\!\!\!\!\!\!\!\!\!\bfsi(\bfx)\!&=&\! \bfsi^I(\bfx)\!+\!\bfsi^{II}(\bfx), \nonumber\\
\bfsi^I(\bfx)&=&\bfsi^I_0(\bfx)\!+\!\bfsi^I_1(\bfx),
\bfsi^{II}(\bfx)= \bfsi^{II}_0(\bfx)\!+\!\bfsi^{II}_1(\bfx),
\EEEQ
with the sources
\BBEQ
\label{3.11}
\bfb^I(\bfx)&=&\bfb(\bfx),\ \ {\bfga}^I({\bf x})={\bf 0}, \\
\label{3.12}
\bfb^{II}(\bfx)&=&{\bf 0},\ \ {\bfga}^{II}({\bf x})={\bfga}({\bf x}).
\EEEQ
The decomposition fields (\ref{3.8})–(\ref{3.10}) are commonly employed (see, e.g., \cite{Buryachenko`2022}), typically in conjunction with homogeneous displacement boundary conditions:
\BB
\label{3.13}
\bfu^I(\bfx)=\bfep^{w\Gamma}\cdot \bfx,\ \ \bfu^{II}(\bfx)={\bf 0},
\EE
for $\bfx\in w_{\Gamma}$, with $\bfep^{w_{\Gamma}}=$const.
In contrast to the conventional homogeneous loading (\ref{3.13}), we consider the case of loading generated by both a forcing term (\ref{2.8}) and thermal effects (\ref{2.9}) of compact support. Under such conditions, (\ref{3.12}) is substituted by the more general decomposition (\ref{3.8})–(\ref{3.10}). The fields $\bfep^I_0(\bfx)$, $\bfu^I_0(\bfx)$, $\bfsi^I_0(\bfx)$ and $\bfep^{II}_0(\bfx)$, $\bfu^{II}_0(\bfx)$, $\bfsi^{II}_0(\bfx)$ then represent the corresponding responses in the homogeneous infinite matrix subjected to {\color{black} loadings (\ref{3.11}) and (\ref{3.12}), respectively. The total fields 
$\bfu(\bfx), \bfep(\bfx)$, and $\bfsi(\bfx)$ are decomposed on the fields $^{(I)}$ and $^{(II)}$ corresponding to the loading (\ref{3.11}) and (\ref{3.12}), respectively.}

Let us examine the case of a single inclusion $v_i$ embedded in an infinite homogeneous matrix.
The displacement field $\bfu^I$ can be characterized by the pair of equations (with $\bfu^I_0=\bfu^{b(0)}(\bfx)$):
\BBEQ
\label{3.14}
{\bfcL}^{(0)}(\bfu^I_0)
(\bfx) &=& -\bfb(\bfx),\\
\label{3.15}
{\bfcL}(\bfu)
(\bfx)
&=& {\bfcL}^{(0)}(\bfu^I_0).
\EEEQ
Equation (\ref{3.14}) defines the reference displacement field in the infinite matrix under the action of the body force $\bfb(\bfx)$, while (\ref{3.15}) establishes the relation between the actual operator $\bfcL$ and its homogeneous counterpart ${\bfcL}^{(0)}$.
Substitution of these relations into the equilibrium equation (\ref{2.1}) gives the differential form
\BB
\label{3.16}
{\color{black}\nabla\cdot}\bfL^{(0)}\nabla \bfu_1^I(\bfx)=- {\color{black}\nabla\cdot}\bfL_1(\bfx)\nabla\bfu(\bfx),
\EE
which describes the perturbation of the displacement field due to the local inhomogeneity of material properties.
Equation (\ref{3.16}) can be recast in an implicit integral form, leading to the following representation for the strain field:
\BB
\label{3.17}
\bfep^I(\bfx) =\bfep^{b(0)}(\bfx)+\int\bfU^{(0)}(\bfx-\bfy){\color{black}\bfL_1}(\bfy)\bfep^I(\bfy)V_i(\bfy)~d\bfy.
\EE
Here, $\bfep^{b(0)}(\bfx)$ denotes the background strain field in the homogeneous matrix, $\bfU^{(0)}$ is the Green-type kernel of the reference medium, and the integral term accounts for the interaction between the inclusion $v_i$ and the surrounding matrix.

For the evaluation of the residual fields ${(\cdot)}^{II}$, the system of equations analogous to (\ref{3.14})–(\ref{3.15}) must be replaced by the following relations, corresponding to the thermal loading problem ($\bfep^{II}(\bfx)=\bfep^{\theta}(\bfx)$):
\BBEQ
{\bfcL}^{(0)}(\bfu^{II}_0,\bfal^{(0)})
(\bfx) &=& {\bf 0},\nonumber \\
\label{3.18}
{\bfcL}^{(0)} (\bfu_1^{II},{\bf 0}))&=&-\bfcL_1(\bfu^{II}, \bfal_1)(\bfx).
\EEEQ
The first equation defines the homogeneous reference problem in the infinite matrix with constant thermal parameter $\bfal^{(0)}$, while the second equation governs the perturbation field induced by the inhomogeneous part $\bfal_1$ and the operator $\bfcL_1$.
The right-hand side of (\ref{3.18}$_2$) can be interpreted as the action of a fictitious body force, yielding the relation
\BB
\label{3.19}
\nabla\bfL^{(0)}\nabla \bfu_1^{II}(\bfx)=- \nabla(\bfL_1(\bfx)\nabla\bfu^{II}(\bfx)+\bfal_1(\bfx)).
\EE
This representation makes it possible to apply the Green’s function formalism.
By introducing the displacement Green’s tensor $\bfG^{(0)}$ of the homogeneous reference medium, Eq. (\ref{3.19}) is reduced to the integral equation
\BB
\label{3.20}
\bfu^{II}_1(\bfx) =\bfG^{(0)}*{\nabla}(\bfL_1\bfep^{II}+\bfal_1).
\EE
A symmetrized differentiation of (\ref{3.20}), followed by integration by parts and the use of the identity $\nabla_{\bf y} = -\nabla_{\bf x}$, leads to the following integral representation for the strain field:
\BB
\label{3.21}
\bfep^{II}(\bfx) =\bfep^{\theta}(\bfx)+\int\bfU^{(0)}(\bfx-\bfy)[\bfL_1(\bfy)\bfep^{II}(\bfy)+\bfal_1(\bfy)]V_i(\bfy)~d\bfy.
\EE
Here, $\bfep^{\theta}(\bfx)$ is the background thermal strain in the homogeneous reference medium, $\bfU^{(0)}$ denotes the symmetrized derivative of the Green’s tensor (i.e., the strain–strain kernel), and the integral term represents the effect of the inclusion $v_i$ through the contrast in elastic moduli $\bfL_1$ and thermal parameters $\bfal_1$.
It is important to note that the derivation of (\ref{3.21}) parallels the classical procedure for constant thermal parameters $\bfal^{(0)}$ (corresponding to the case of uniform temperature change $\theta \equiv \mathrm{const}$, see, e.g., \cite{Buryachenko`2007}). However, the present formulation explicitly extends the result to the more general and practically relevant situation of temperature change with compact support as introduced in (\ref{2.9}).

By summing Eqs. (\ref{3.17}) and (\ref{3.21}) in the same manner as in Eq. (\ref{3.8}$_1$), we obtain the total representation for the strain field inside the inclusion $v_i$:
\BB
\label{3.22}
\bfep(\bfx) =\bfep^{b(0)}(\bfx) +\bfep^{\theta}(\bfx)+\int\bfU^{(0)}(\bfx-\bfy)[\bfL_1(\bfy)\bfep(\bfy)+\bfal_1(\bfy)]V_i(\bfy)~d\bfy,
\EE
where $\bfep^{b(0)}(\bfx)$ denotes the strain field generated in the homogeneous reference medium by the body force (\ref{2.8}), $\bfep^{\theta}(\bfx)$ corresponds to the thermal strain induced by the temperature change (\ref{2.9}), and the integral term accounts for the perturbation due to the material inhomogeneity inside the inclusion $v_i$.
Equation (\ref{3.22}) thus provides a compact but general integral formulation for the strain field in the presence of both mechanical and thermal sources. {\color{black} This representation (\ref{3.22}) gives the stress equation}
\BB
\label{3.23}
\bfsi(\bfx) =\bfsi^{b(0)}(\bfx) +\bfsi^{\theta}(\bfx)+\int{\bf \Gamma}^{(0)}(\bfx-\bfy)[\bfM_1(\bfy)\bfsi(\bfy)+\bfbe_1(\bfy)]V_i(\bfy)~d\bfy,
\EE
where $\bfsi^{b(0)}(\bfx)$ and $\bfsi^{\theta}(\bfx)$ denote the reference stress fields corresponding to purely mechanical and purely thermal loadings of the infinite matrix, respectively, while ${\bf \Gamma}^{(0)}$ is the stress Green kernel associated with the homogeneous reference medium. The transition from (\ref{3.22}) to (\ref{3.23}) relies on the following tensor identities:
\BB
\label{3.24}
\bfL_1(\bfep- \bfbe) =-\bfL^{(0)}\bfM_1\bfsi\ \
{\rm and} \ \ \bfep = [\bfM^{(0)}\bfsi+\bfbe^{(0)}]+
[\bfM_1\bfsi+\bfbe_1]{\color{black},}
\EE
which express the relation between strains and stresses in terms of the compliance tensors $\bfM^{(0)}$, $\bfM_1$ and the corresponding thermal expansion tensors $\bfbe^{(0)}$, $\bfbe_1$.
Together, (\ref{3.22}) and (\ref{3.23}) provide dual strain–stress formulations of the inclusion problem, suitable for subsequent analysis and numerical implementation.

\subsection{Additive general integral equations}

Equations (\ref{3.23}) and (\ref{3.24}) were derived for a single, fixed inclusion $v_i$ embedded in an infinite homogeneous matrix. 
In the case of a composite material (CM) containing multiple inclusions, the influence of all surrounding inclusions on the fixed inclusion $v_i$ can be systematically accounted for through the Generalized Integral Equation (GIE, {\color{black} \cite{Buryachenko`2025})}). A particular form of this, known as the Additive GIE (AGIE), directly sums the contributions from all neighboring inclusions without applying renormalization procedures.
{\color{black} AGIEs and their detailed solutions are presented in \cite{Buryachenko`2025} for random-structure composite materials described by probability density functions (see Subsection 2.2). Since periodic structures may be regarded as particular cases of the random-structure descriptors (2.20), the subsequent manipulations for periodic composites represent direct extractions from the corresponding developments for random composites. Interested readers are referred to \cite{Buryachenko`2025}, which provides a much more detailed and comprehensive treatment (including 542 references).}

For a point $\bfx$ inside the inclusion $v_i$ of periodic structure CMs, the AGIE reads:
\BBEQ
\label{3.25}
\!\!\!\!\! \bfxi_i (\bfx)&=& \bfxi^{b(0)} ({\bf x})+\bfxi^{\theta}(\bfx)+\sum_{{\bf x}_j\in{\bf \Lambda}}\bfcL_j^{\xi\zeta}(\bfx-\bfx_j,\bfze),
\EEEQ
where the {\color{black} infinite} summation captures how each neighboring inclusion $v_j$ directly alters the field at $\bfx \in v_i$.
The first two terms, $\bfxi^{b(0)} (\bfx)$ and $\bfxi^{\theta} (\bfx)$, represent the deterministic fields generated by the body force $\bfb(\bfx)$ (Eq. (\ref{2.8})) and the compact-support temperature change $\theta(\bfx)$ (Eq. (\ref{2.9})), respectively, in the homogeneous infinite matrix. These serve as the background or reference fields before the perturbations due to the inclusions are applied.
The tensors
\BBEQ
\label{3.26}
{\bfxi} (\bfx)-\overline{\bfxi} (\bfx):=\bfcL_j^{\xi\zeta}(\bfx-\bfx_j, \bfze) =\int \bfU^{\gamma}(\bfx-\bfy) \bfze(\bfy)V_j(\bfy) ~d\bfy
\EEEQ
are called {\it perturbators}, and they represent the incremental perturbation induced at the point $\bfx \in v_i$ by the presence of the inclusion $v_j$ under the action of the effective field $\overline{\bfxi} (\bfx)$. Physically, the perturbators quantify how the inclusion $v_j$ modifies the local field in $v_i$ relative to the averaged or effective field in the matrix.
The superscripts $^{\xi\zeta}$ in $\bfcL^{\xi\zeta}$ indicate the mapping from the variable $\bfze$ on the right-hand side of the integral to the resulting variable $\bfxi$ on the left-hand side. This notation clarifies which type of field (mechanical strain, stress, or thermal strain) is being perturbed and which quantity acts as the source of perturbation.
The term {\it Additive GIE} is chosen by analogy with Additive Manufacturing, reflecting the fact that the perturbations from all surrounding inclusions are directly summed in Eq. (\ref{3.25}) without introducing any additional renormalization or correction terms that are typical in the classical GIE approach (see Subsection 3.4 for a detailed comparison). In this framework, the total field inside $v_i$ is obtained as a superposition of the background field and all inclusion-induced perturbations, providing a transparent and computationally convenient formulation for heterogeneous media.

To the authors’ knowledge, no studies explicitly address inhomogeneous thermal loading in composite materials (CMs), despite its practical relevance in contexts such as laser-based additive manufacturing \cite{Yang`et`2019} and military applications \cite{Isakari`et`2017}. In these cases, laser pulses induce rapid local heating, steep temperature gradients, and high thermal stresses. Conventional analyses often assume a homogeneous medium \cite{Yilbas`2013}, but in CMs, thermal and stress fields vary on the scale of microstructural heterogeneities, highlighting the need for nonlocal formulations like Eqs. (\ref{3.25}) to capture the true thermomechanical response.

Figure 1 provides a schematic illustration of the types of loading with compact support considered in this study. It includes a deterministic, self-equilibrated body force $\bfb(\bfx)$, a thermal change $\theta(\bfx)$, and an inhomogeneous eigenstress distribution $\bfal^{(0)}(\bfx)$ that represents preexisting internal stresses in the matrix. In addition, the term $\bfal_1(\bfx)$ {\color{black} (vanishing in the matrix)} captures the eigenstress mismatch arising from the presence of the surrounding periodic inclusion field, reflecting the perturbations induced by heterogeneities in the composite microstructure. Together, these components define the localized loading environment that drives the subsequent strain and stress responses within the material.

The operators $\bfcL_j^{\xi\zeta}(\bfx - \bfx_j, \bfze)$ (\ref{3.26}) correspond to the fundamental micromechanical problem of a single inclusion embedded in an infinite homogeneous matrix. We now introduce another class of perturbators:
\BB
\label{3.27}
\bfcL^{\xi\xi}_k(\bfx-\bfx_k,\overline{\bfxi }) \equiv \bfxi(\bfx)-\overline{\bfxi}(\bfx),
\EE
which represent explicit solutions of problems of the type given in (\ref{3.26}).
More precisely, the perturbators $\bfcL^{\xi\zeta}_i(\bfx-\bfx_k,\bfze)$ should be regarded merely as symbolic notations for a problem to be solved, expressed through
$\bfxi - \overline{\bfxi} = \bfcL^{\xi\zeta}_i(\bfx-\bfx_k,\bfze)$ (\ref{3.26}$_1$),
whereas the perturbators $\bfcL^{\xi\xi}_i(\bfx-\bfx_k ,\overline{\bfxi})$ denote the actual solutions to this problem.

Equation (\ref{3.25}) is reformulated in terms of effective fields ($\bfx\in v_i$)
\BBEQ
\label{3.28}
\!\!\!\!\! \overline {\bfxi}^{[n+1]} _i (\bfx)&=& \bfxi^{b(0)} ({\bf x})+\bfxi^{\theta}(\bfx)+{\sum_{{\bf x}_j\in{\bf \Lambda}}}^\prime\bfcL_j^{\xi\xi}(\bfx-\bfx_j,\overline{\bfxi}^{[n]} ).
\EEEQ
Here $\sum'$ indicates exclusion of the self-term $v_j=v_i$, since neighboring inclusions $v_j$ directly influence the field at $\bfx \in v_i$. The initial approximation is defined by the driving term (zero-order), $ \overline{\bfxi}^{[0]}(\bfx)=\bfxi^{b(0)} ({\bf x})+\bfxi^{\theta}(\bfx)$,
as given in Eqs.~(\ref{3.6})–(\ref{3.8}).
The iterative scheme (\ref{3.28}) then generates a Neumann series, yielding the limit solution for $\bfx\in v_i$:
\BB
\label{3.29}
\overline{\bfxi}_i(\bfx) := \lim_{n\to \infty}\overline{\bfxi}^{[n+1]}_i(\bfx) =\widehat{\bfcD}_i^{\xi b\theta}(\bfb,\theta,\bfx).
\EE
This leads to the formulation of the statistical averages of the field within the fixed inclusion, specifically the conditional averages for
$\bfx\in v_i$ ($\bfxi=\bfep$)
\BBEQ
\label{3.30}
\!\!\!\!\!\!\!\!\!\!{\bfxi}_i(\bfx) \!\!&=&\!\! \widehat{\bfcD}_i^{\xi b\theta}(\bfb,\theta,\bfx) +
\bfcL_i^{\xi\xi}(\bfx-\bfx_i, \widehat{\bfcD}_i^{\xi b\theta}(\bfb,\theta,\bfx)),\\
\label{3.31}
\!\!\!\!\!\!\!\!\!\!\!\!{\bfsi}_i(\bfx) \!\!&=&\bfL(\bfx)\bfep_i(\bfx).
\EEEQ
The tensor $\widehat{\bfcD}_i^{\xi b\theta}$ (\ref{3.29}) is an inhomogeneous function of the coordinates of the fixed inclusion,
reflecting interactions with all intersecting inclusions, whereas in the dilute case it reduces to 
$\bfI(\bfxi^{b(0)}(\bfx)+\bfxi^{\theta}(\bfx))$.

\vspace{1.mm} \noindent \hspace{30mm}
\parbox{8.8cm}{
\centering \epsfig{figure=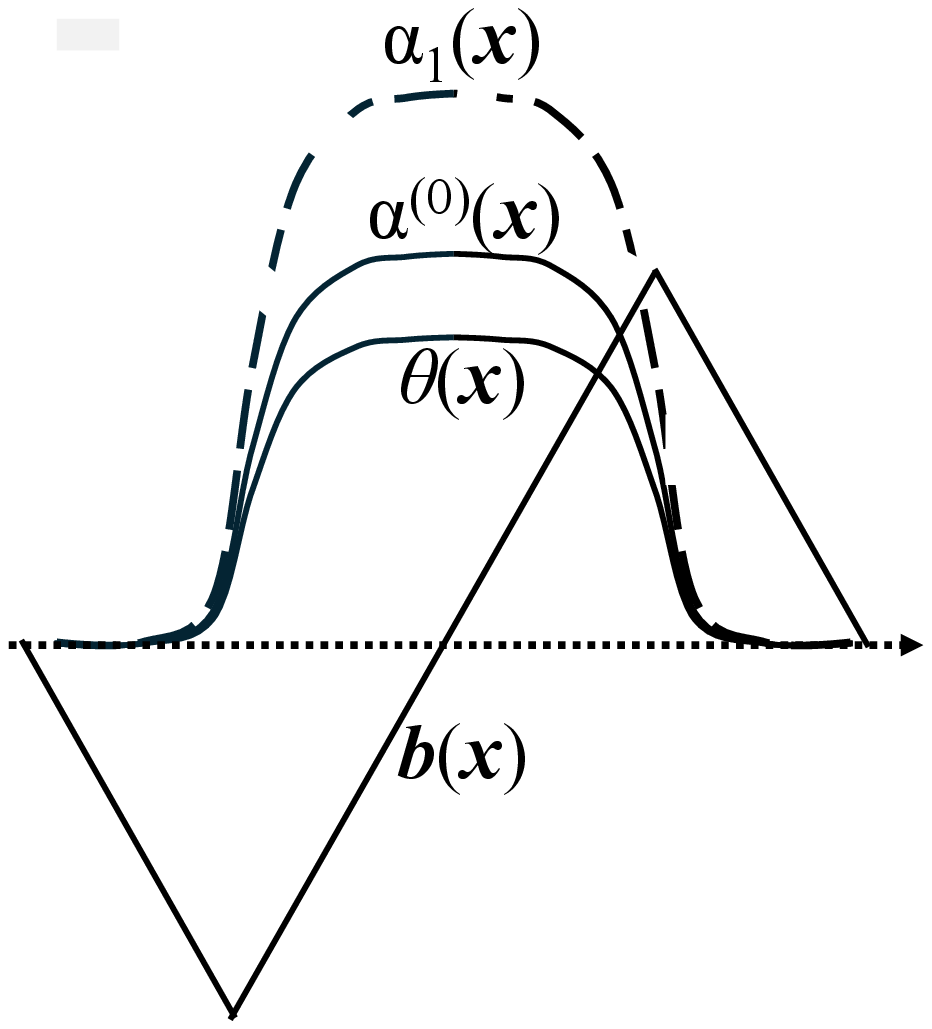, width=7.8cm}\\ \vspace{-50.mm}
\vspace{132.mm}
\vspace{-117.mm} \tenrm \baselineskip=8pt
{{ Fig. 2:} Continuous $\bfb(\bfx),\ \theta(\bfx),\ \bfal^{(0)}(\bfx)$; discontinuous periodic $\bfal_1(\bfx)$}}
\vspace{0.mm}

\subsection{General integral equations}
A central notion in analytical micromechanics is the GIE, which establishes an exact relation between random fields at a point and those in its neighborhood. {\color{black} Althogh GIEs are nor directly required for the main approach of this paper, we consider it because GIEs are closely related with AGIE and have the same sourse}. 

For periodic composites, different formulations of GIE—ordered by their level of generality—are presented in \cite{Buryachenko`2015}, summarized in \cite{Buryachenko`2022} ($\bfx\in w$), and recast in therms {\color{black} of Eq, (\ref{3.26}) $(\bfxi=\bfep,\ \bfze=\bftau)$:}
\BBEQ
\label{3.32}
\!\!\!\!\!\!\!{\bfep}({\bf x}) &=& {\bfep}^{w \Gamma}
+\int{\bfU}^{(0)}(\bfx-\bfy)[\bfL_1(\bfy)\bfep(\bfy)+\bfal_1(\bfy)]~d{\bf y},
\\ 
\label{3.33}
\!\!\!\!\!\!\!{\bfep}({\bf x})&=&\langle {\bfep}\rangle 
+\int \bfcL^{\xi\tau}(\bfx-\bfx_j,\bftau\delta(\bfx-\bfx_j|;\bfx_i)-{\bfU}^{(0)}(\bfx-\bfx_j)
\lle[\bfL_1(\bfy)\bfep(\bfy)+\bfal_1(\bfy)]\rle]~d{\bf x}_j,\\
\label{3.34}
\!\!\!\!\!\!{\bfep}({\bf x})
&=&
\langle {\bfep}\rangle_{\bf x} ({\bf x})
+\int\Big[\bfcL^{\xi\tau}(\bfx-\bfx_j,\bftau)\delta(\bfx-\bfx_j|;\bfx_i)
-\lle \bfcL^{\xi\tau}(\bfx-\bfx_j,\bftau)\rle_{\bf x} \Big]
~d{\bf x}_j.
\EEEQ
Hereafter $\lle(\cdot)\rle$ and $\lle(\cdot)\rle(\bfx)$ are the statistical averages whreas $\lle(\cdot)\rle_{\bf x}$ is the average over the moving window (\ref{2.19}); $\delta(\bfx-\bfx_j|;\bfx_i)=\sum^{\prime}_j\delta(\bfx-\bfx_j)$. 

The progression of the Generalized Integral Equation (GIE) from Eq.(\ref{3.32}) to Eq. (\ref{3.34})—with key contributions from \cite{{Rayleigh`1892},{Shermergor`1977},{Khoroshun`1978},{OBrian`1979}}—is thoroughly examined in \cite{Buryachenko`2022}. Importantly, the integral in Eq.~(\ref{3.32}) lacks absolute convergence, unlike Eq.~(\ref{3.33}), first formulated by Rayleigh \cite{Rayleigh`1892} under the assumptions $\lle\bfep\rle(\bfx)\equiv {\rm const}$ and $\bfal_1(\bfy)\equiv \mathbf{0}$. This latter form resolves issues with the asymptotic decay of the generalized Green operator $\bfU$ at infinity (scaling as $|\bfx-\bfy|^{1-d}$), and removes any need to specify the geometry of the integration domain $w$.
As a result, methods such as regularization, renormalization, or auxiliary mixed-boundary problems—commonly invoked to manage divergence at infinity—become unnecessary (see \cite{Buryachenko`2022} for references). The shift in the statistical averaging operator from Eq.~(\ref{3.33}) to Eq.~(\ref{3.34}) marks the emergence of a new paradigm in micromechanics, namely Computational Analytical Micromechanics (CAM) \cite{Buryachenko`2022}. Furthermore, \cite{Buryachenko`2022} demonstrates that the exact Eq.~(\ref{3.34}) reduces to the approximate Eq.~(\ref{3.33}) when the EFH assumption holds. In this sense, CAM represents a major conceptual advance, building upon Rayleigh’s original intuitive formulation of the GIE framework \cite{Rayleigh`1892} (see for details Chapter 7 in \cite{Buryachenko`2022}).

A {\it centering procedure} is applied to Eqs. (\ref{3.25}) by subtracting their statistical averages from both sides, yielding the more general Generalized Integral Equation (GIE) in Eq.(\ref{3.34}). Unlike Eqs.(\ref{3.25}), which are restricted to the particular loading BFCS (\ref{2.8}) and TCDC (\ref{2.9}), the centered form in Eq.(\ref{3.34}) remains valid for arbitrary inhomogeneous loading fields $\langle \bfxi \rangle(\bfx)$. A central benefit of this centering step is the appearance of the renormalizing term
$\lle \bfcL^{\theta\zeta}(\bfx-\bfx_j,\bfze)\rle_{\bf x}$,
which guarantees absolute convergence of the integral expressions.
What appears to be a straightforward reduction—from Eq.(\ref{3.32}) to Eq.(\ref{3.33}), and further to Eq.(\ref{3.34})—actually brings a deeper, less obvious advantage. Specifically, Eqs.(\ref{3.33}) and (\ref{3.34}) were previously obtained {\color{black} in} Chapter 7 \cite{Buryachenko`2022} under the simplifying assumption $\theta(\bfx)\equiv {\rm const.}$, where, for aligned identical inclusions, the polarization term $\bfal_1(\bfy)$ is translation-invariant, even in a functionally graded medium (FGM). Remarkably, however, the same integral equations remain valid when $\theta(\bfx)\not\equiv {\rm const.}$—as in the case of TCDC (\ref{2.9})—where periodic $\bfal_1(\bfy)$ loses its translation invariance, even for statistically homogeneous fields with aligned inclusions. {\color{black} Thus, in this paragraph—without introducing any additional equations—we arrived at a fundamentally new result: Eq. (\ref{3.34}) is valid for any non-constant function $\theta(\bfx)\not\equiv $const., although previously its correctness had been established only for the case $\theta(\bfx)\equiv $const}. 

{\color{black} It is remarkable that both GIEs and AGIEs (for composites of random and periodic structures, see Chapter 7 in \cite{Buryachenko`2022} and \cite{Buryachenko`2022}, respectively) follow from the same, seemingly well-known but actually incorrect equation (\ref{3.32}). 
This common source (\ref{3.32}) clearly and simultaneously highlights both the fundamental differences between GIEs and AGIEs and the immediate reduction of AGIEs to GIEs.
Indeed,} two distinct developmental trajectories have emerged in addressing the shortcomings of the original Eq.~(\ref{3.32}), each giving rise to a fundamentally new direction in micromechanics. The first, initiated by Lord Rayleigh \cite{Rayleigh`1892}, involved a pivotal modification of the integral term in Eq.~(\ref{3.32}).
This conceptual innovation laid the foundation for more than a century of progress, culminating in the evolution from Eq.~(\ref{3.33}) to the generalized formulation in Eq.~(\ref{3.34}). The latter represents the culmination of this line of inquiry and may be regarded as the ``second background" (called also CAM) of micromechanics —the most significant advance in the field since Rayleigh’s seminal work \cite{Rayleigh`1892} ({\color{black} this statement is proved in Chapter 7 in \cite{Buryachenko`2022})}. 
By contrast, the second trajectory preserves the original integral structure of Eq.(\ref{3.32}) but introduces a decisive conceptual innovation: {\color{black} the substitution of the classical free term $\bfep^{w\Gamma}\equiv$const.
with the generalized fields $\bfep^{b(0)}(\bfx)\not\equiv$const. and $\bfep^{\theta}(\bfx)
\not\equiv$const.} produced by compactly supported loading (\ref{2.8}) and (\ref{2.9}). This transformation marks a shift from conventional homogeneous boundary conditions (\ref{2.13}) to the broader framework of BFCS (\ref{2.8}) and TCCS
(\ref{2.9}). The resulting formulation, embodied in Eq.(\ref{3.25}), establishes the foundation of what is referred to herein as the new philosophy of micromechanics (see Conclusion). Taken together, these two pathways do not merely reflect technical adjustments but rather mark profound conceptual shifts. 
{\color {black} The first pathway (including GIE (\ref{3.34})) culminates in the most impactful advance since Rayleigh’s seminal GIE formulation. In contrast, the second pathway, initiated by AGIEs, embodies a fundamental reconceptualization of micromechanical theory—arguably the most profound transformation since
the (first) classical background of Poisson, Faraday, Mossotti, Clausius, Lorenz, and Maxwell (1824–1879) based on the EFH.
AGIE is independent of GIE in both formulation and interpretation. Equation (\ref{3.25}), together with its solutions (\ref{3.28})–(\ref{3.31}) for periodic and random composites, is completely detached from the EFH and therefore constitutes a new philosophy of micromechanics
(see Eq. (\ref{6.1}) with subsequent comments and \cite{Buryachenko`2025} for a detailed explanation why AGIEs form a new phylosophy of micromechanics). 
Furthermore, the {\it centering procedure} described above provides an immediate reduction of AGIEs to GIEs that are applicable to arbitrary loadings—not limited to the CS loadings (\ref{2.8}) and (\ref{2.9}) -- and to general inhomogeneous remote boundary conditions, instead of only the homogeneous condition (2.13); see \cite{Buryachenko`2025}.
}

It is worth noting that alternative GIEs exist for periodic composite materials, namely the classical Lippmann–Schwinger (L-S) equation (\ref{3.35}) and its modified counterpart (\ref{3.36}) (with $\theta \equiv {\rm const.}$):
\BBEQ
\label{3.35}
\!\!\!\!\!\!{\bfep}({\bf x})
&=&
\langle {\bfep}\rangle ({\bf x})
+\int{\bfU}^{\rm p}(\bfx-\bfx_j) [\bfL_1(\bfy)\bfep(\bfy)+\bfal_1(\bfy)]d{\bf y},
\\ 
\label{3.36}
\!\!\!\!\!\!{\bfep}({\bf x})
&=&
\bfep^{b(0)}+\bfep^{\theta}(\bfx)
+\int{\bfU}^{\rm p}(\bfx-\bfx_j) [\bfL_1(\bfy)\bfep(\bfy)+\bfal_1(\bfy)]d{\bf y},
\EEEQ
where the free term {\color{black}$\bfep^{b(0)}+\bfep^{\theta}(\bfx)$, and the Green function ${\bfU}^{\rm p}$ are periodic with respect to $\Omega^b_{00}$ and $\Omega_{00}$}, respectively. This periodicity of Eqs. (\ref{3.35}) and (\ref{3.36}) enables the use of FFT-based solvers.
FFT-based numerical homogenization methods, pioneered by Moulinec and Suquet \cite{{Moulinec`S`1994}, {Moulinec`S`1998}}, provide a computationally efficient alternative to FEM, reducing the complexity from $O(N^2)$ to $O(N \log N)$ by exploiting the discrete Fourier transform (DFT). For comprehensive overviews, see \cite{{Lucarini`et`2022}, {Schneider`2021},{Segurado`et`2018},{Zeman`et`2010}}. More recently, Buryachenko \cite{Buryachenko`2025b} has introduced several FFT algorithms for solving Eq. (\ref{3.36}) under BFCS loading conditions (\ref{2.8}).

{\bf Comment 3.1.} The approach (\ref{3.1})–(\ref{3.31}) was originally formulated for a finite number of inclusions embedded in an infinite matrix. Importantly, the periodicity (\ref{2.11}) of the compact support (CS) loading (\ref{2.8}) and (\ref{2.9}) does not enter into Eqs. (\ref{3.1})–(\ref{3.31}), which could, in principle, be solved by alternative techniques such as FEA—the essential requirement being only the compact support conditions (\ref{2.8}) and (\ref{2.9}). The periodicity assumption for CS loading (\ref{2.10}) is introduced solely to enable the solution of Eq. (\ref{3.36}) (the analogue of Eq. (\ref{3.25})) via the FFT method (see \cite{Buryachenko`2025b} for details). Consequently, the DNS solution of Eq. (\ref{2.5}) with CS loading (\ref{2.8}) and (\ref{2.9}) is regarded as available for both the systems with a finite number of inclusions in an infinite matrix {\color{black} (or a finite specimen with traction-free boundaries, the problem $\bfcP^{\rm iso}$ considered in Subsection 2.1) and for periodic configurations satisfying (\ref{2.10}) and (\ref{2.14})
(the problem $\bfcP^{\rm per}$ considered in Subsection 2.1). These DNS solutions are }then employed in Section 4, specifically in Eqs. (\ref{4.5}) and (\ref{4.6}).

\section{Classical and new RVE concepts}
\setcounter{equation}{0}
\renewcommand{\theequation}{4.\arabic{equation}}

The concept of the Representative Volume Element (RVE), first introduced by \cite{Hill`1963}, has a long and nuanced history, often accompanied by debate and varying interpretations (see \cite{Ostoja`et`2016} for a detailed discussion). To preserve the original intent and highlight the foundational significance of Hill’s definition, we refer directly to his seminal work \cite{Hill`1963}, which establishes a rigorous framework for the RVE concept under the condition $\bfb(\bfx)\equiv {\bf 0}$.

\noindent{\bf Definition 4.1.} {\it Representative volume element (RVE)
(a) is structurally entirely typical of the whole mixture on average, and
(b) contains a sufficient number of inclusions for the apparent overall moduli to
be effectively independent of the surface values of traction and displacement, so
long as these values are ‘macroscopically uniform'.... The contribution of this surface layer to any average can be negligible by taking the sample large enough.}

The main purpose of the RVE is to enable the determination of the effective moduli $\bfL^*$ through the calculation of phase-averaged field quantities. In the case of composite materials with periodic microstructures, the RVE is identical to the Unit Cell (RVE~$\equiv$~UC), and the imposition of remote homogeneous boundary conditions (\ref{2.13}$_1$) naturally reduces to the application of periodic boundary conditions (PBC, (\ref{2.11})). 
A representative response can be obtained via DNS of microstructural volume elements (MVEs), either synthetically generated or experimentally extracted (e.g., by micro-CT). Homogenized properties $\bfL^{\rm A}_{\rm KUBC}$ and $\bfM^{\rm A}_{\rm SUBC}$ are computed under kinematically (\ref{2.13}$_1$) and statically (\ref{2.13}$_2$) uniform boundary conditions, respectively. Their difference
\BB
\label{4.1}
\bfL^{\rm A}_{\rm KUBC}-(\bfM^{\rm A}_{\rm SUBC})^{-1}\to {\bf 0},
\EE
decreases with increasing MVE size, enabling estimation of the RVE size and effective moduli $\bfL^*$. While increasing the number of realizations can improve statistical convergence, it does not remove edge-related scale effects. In practice, intuitive RVEs are constructed either from simulated random inclusion fields or from micro-CT-based image models (see, e.g., \cite{Konig`et`1991, Ohser`M`2000, Torquato`2002}). A schematic illustration of the RVE concept is shown in Fig. 2, depicting a micro-CT scan under remote homogeneous loading (\ref{2.13}).

\vspace{1.mm} \noindent \hspace{30mm}
\parbox{8.8cm}{`
\centering \epsfig{figure=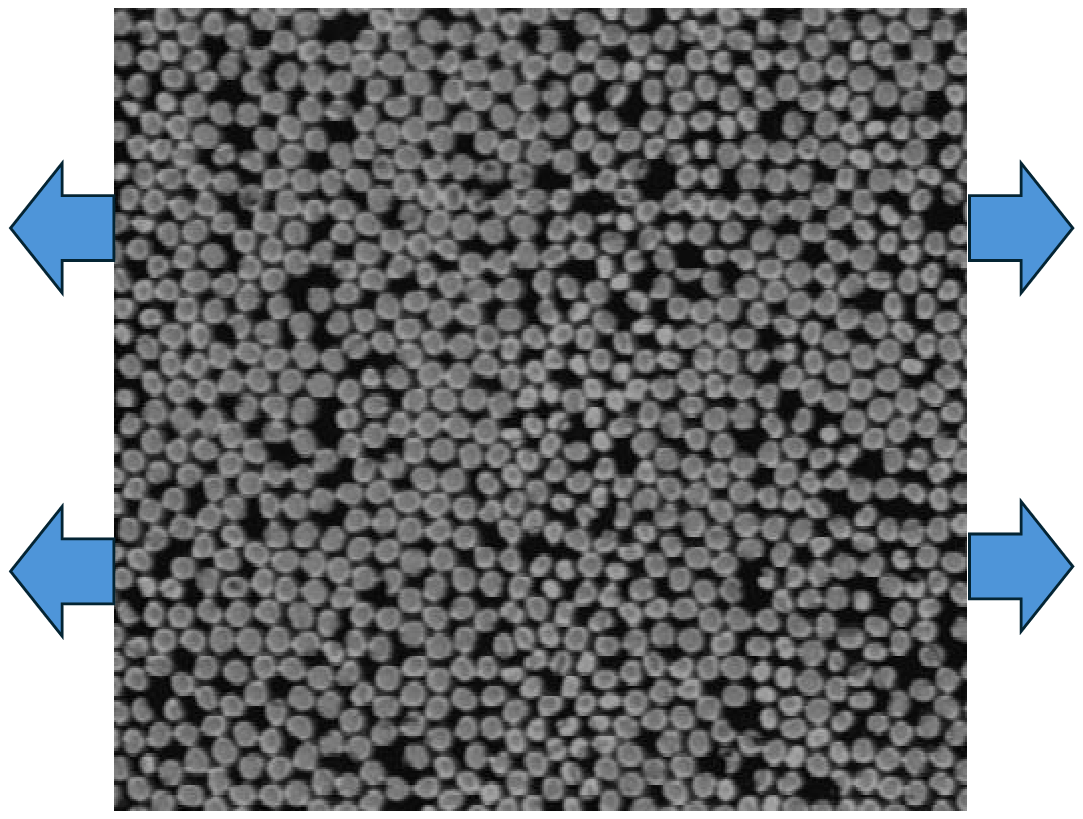, width=7.8cm}\\ \vspace{-50.mm}
\vspace{122.mm}
\vspace{-117.mm} \tenrm \baselineskip=8pt
{\color{black} { Fig. 3:} CT image with remote BC (\ref{2.13}) }}
\vspace{2.mm}

Under PBC (\ref{2.15}) (or (\ref{2.13})), the effective moduli $\bfL^*$ (``upscaling") and the field concentration factors ${\color{black} \bfA^*}(\bfx)$ for points $\bfx \in v_i$ (``downscaling") are given by
\BBEQ
\!\bfL^*=\bfL^{(0)}+\bfR^*, \ \ \lle\bftau\rle=\bfR^*\lle\bfep\rle,\ \ 
\label{4.2}
\lle\bfep\rle_i(\bfx)=\!\bfA^*(\bfx)\lle \bfep\rle
\EEEQ
emphasizing the intrinsic link between upscaling and downscaling in the homogenization framework {\color{black}(as two sides of the same coin)}.

Under {\color{black} PBC (\ref{2.15}), Eqs. (\ref{2.13})} guarantee that both the effective moduli $\bfL^*$ and the strain field $\bfep(\bfx)$ are independent of the chosen unit cell $\Omega_{ij}$. Problems of this type fall naturally within the scope of classical homogenization theories for periodic structures, including {\it asymptotic expansions} and {\it computational homogenization} (see Introduction).
This elegant invariance collapses, however, once BFUC loading $\Omega_{00}^b$ (\ref{2.8}) is applied in conjunction with the CTCS (\ref{2.9}), since condition (\ref{2.15}) is violated. In such a setting, the strain field $\bfep(\bfX)$ (with $\bfX \in \Omega_{00}^b$) acquires an explicit dependence on the specific unit cell $\Omega_{ij}^b$ containing the point $\bfX$. The striking consequence is that the powerful arsenal of homogenization methods {\color{black}(such as {\it asymptotic expansions} and {\it computational homogenization})}, long regarded as indispensable for periodic media, {\color{black} not only becomes superfluous but actually ceases to be applicable for such problems}.

To remove this dependency, we study periodic CMs under a fixed body force while allowing the unit-cell grid $\bfx_{\bf m} \in \bfLa$ to undergo a rigid translation. For linear elasticity (\ref{2.5}), the strain solution is denoted $\widetilde\bfep_0(\bfx)$. When the grid shifts by $\bfchi$ ($\bfLa_0 \to \bfLa_{\bf\chi}$), the stiffness $\bfL(\bfx,\bfchi)$, phase indicator $V_i(\bfx,\bfchi)$, and strain field $\bfep(\bfx,\bfchi)$ transform accordingly
\BBEQ
\label{4.3}
\bfL(\bfx,\bfchi)&=&\bfL_0(\bfx-\bfchi), \ \ V_i(\bfx,\bfchi)=V_{i0}(\bfx-\bfchi),\\
\bfb(\bfx,\bfchi)&=&\bfb_0(\bfx),\ \ \theta(\bfx,\bfchi)=\theta_0(\bfx), \ \ \bfep(\bfx,\bfchi)\not\equiv \widetilde\bfep_0(\bfx-\bfchi).
\label{4.4}
\EEEQ
The validity of Eq.~(\ref{4.4}$_3$) stems from the fixed nature of the loading fields $\bfb(\bfx)$ and $\theta(\bfx)$.

From this point forward, we assume that for each $\bfchi \in \cV_{\bf x}$, the solution of Eq. (\ref{2.5}) for a periodic composite material (\ref{2.14}) under CS loading conditions (\ref{2.8}) and (\ref{2.9}) is available. Practically, the relevant computational domain is a finite-size region enclosing $\bfcB^b$. Such solutions can be obtained by various approaches, including FEM or the volume integral equation method (VIEM, see Subsection 3.3 and Comment 3.1). The periodicity of CS loading (\ref{2.10}) is introduced primarily to facilitate the use of the FFT method for solving the modified L-S equation (\ref{3.36}) (the analogue of Eq. (\ref{3.25})). Consequently, for each translation $\bfchi \in \cV_{\bf x}$, the corresponding solution $\bfep(\bfx,\bfchi)$ from Eq. (\ref{4.4}) defines the effective (macroscopic) fields over $w$
\BBEQ
\label{4.5}
\!\!\!\!\!\!\!\!\!\!\!\!\!\!\!\!\!\!\!\!\!\!\lle\bfep\rle(\bfx)\!&=&\!{1\over\overline
{\cV}_{\bf x}}\int_{{\cal V}_{\rm \bf x}}\bfep(\bfx,\bfchi)~d\bfchi, \ 
\lle\bfep\rle^{(1)}(\bfx)= {1\over\overline{\cV}_{\bf x}}\int_{{\cal V}_{\rm \bf x}}\bfep(\bfx,\bfchi)V_i(\bfx,\bfchi)~d\bfchi, \\
\label{4.6}
\!\!\!\!\!\!\!\!\!\!\!\!\!\!\!\!\!\!\!\!\!\!\!\!\lle\bfsi\rle(\bfx)\!&=&\! {1\over\overline
{\cV}_{\bf x}}\int_{{\cal V}_{\rm \bf x}}\bfsi(\bfx,\bfchi)~d\bfchi,\
\lle\bfsi\rle^{(1)}(\bfx)= {1\over\overline
{{\cal V}}_{\bf x}}\int_{\cV_{\rm \bf x}}\bfsi(\bfx,\bfchi)V_i(\bfx,\bfchi)~d\bfchi,
\EEEQ
Although Eqs.~(\ref{4.5}$_1$) and (\ref{4.6}$_1$) resemble moving-window averages over $\cV_{\bf x}$, they are in fact ensemble averages taken over multiple strain $\bfep(\bfx,\bfchi)$ and stress $\bfsi(\bfx,\bfchi)$ realizations generated by parallel translations of $\bfL(\bfx,\bfchi)$ (\ref{4.4}), rather than by averaging a single solution $\widetilde\bfep_0(\bfx)$. Here, the translations $\bfchi$ are uniformly distributed within the periodicity cell $\cV_{\bf x}$, yielding a general translation-based averaging framework applicable to periodic composites with arbitrary constitutive laws and inhomogeneous loadings
{\color{black} (e.g. (\ref{2.8}) and (\ref{2.9}))}. Each realization can be computed by standard numerical methods (e.g., FEA, FFT). A special case of this scheme—Eq.(\ref{4.6}$_1$)—was first noted in asymptotic homogenization by \cite{Smyshlyaev`C`2000}, based on J.R. Willis’s insight, and later revisited by \cite{Ameen`et`2018}. The statistical averages (\ref{4.5}$_2$) and (\ref{4.6}$_2$) may also be viewed probabilistically, akin to the classical “random coin drop” problem, where inclusions $v_i$ lie on a periodically shifted grid $\bfLa{\bf\chi}$, highlighting the ensemble character of the averaging.
Figure 3 illustrates the averaging scheme of Eq.~(\ref{4.6}) in 1D. A fixed macroscopic point $\bfX \in b({\bf0},B^b)$ corresponds to local points $\bfx_j=\bfX-\bfchi$ within translated grids $\bfLa_j$ ($j=1,\dots,6$). The average stress $\langle \bfsi \rangle(\bfX)$ is obtained by summing $\bfsi(\bfx_j)$ over all realizations, whereas the inclusion average $\langle \bfsi \rangle^{(1)}(\bfX)$ sums only those cases where $\bfX$ lies inside an inclusion (here $j=1,2,3,6$). Owing to nonlocal effects, nonzero stresses $\langle \bfsi \rangle(\bfY)\ne\mathbf{0}$ may occur at points $\bfY \notin b({\bf0},B^b)$ with $\bfb(\bfY)=\mathbf{0}$, again averaged over all grids ($j=1,\dots,6$), while $\langle \bfsi \rangle^{(1)}(\bfY)$ includes only realizations $j=1,2,6$.

Following the averaging procedures (\ref{4.5}) and (\ref{4.6}) with respect to the grid translation $\bfLa{\bf \chi}$, we introduce a dataset $\bfcD^{T}=({\bfcD^I,\bfcD^{II}})$ (called effective dataset) for periodic-structured composite materials. 
Namely, for general BFCS (\ref{2.8}) and TCCS (\ref{2.9}) loadings, we establish formal relations for both macroscopic and microscopic fields in periodic composite microstructures. The total fields $({\cdot})^T$ are decomposed as $({\cdot})^T=(({\cdot})^I,({\cdot})^{II})$, consistent with Eqs. (\ref{3.8})–(\ref{3.10}). Particular attention is given to the macroscopic strain $\langle{\bfep}^T\rangle(\bfx)$ and stress $\langle{\bfsi}^T\rangle(\bfx)$, together with their microscopic counterparts $\langle{\bfep}^T\rangle_i(\bfx)$ and $\langle\bfsi^T\rangle_i(\bfx)$, where $i$ denotes inclusion $v_i$. To enable numerical implementation, an effective dataset is assembled from multiple realizations of compact-support loadings (\ref{2.8}), (\ref{2.9}) with $\bfx \in \mathbb{R}^d$
\BBEQ
\!\!\!\!\!\!\!\!\!\!\!{\bfcD}^{I}&\!=\!&\{\bfcD^{I}_k\}_{k=1}^N, \ \ \ {\bfcD}^{I}_k=\{
\lle{\bfep}^I_k\rle(\bfb_k,0,\bfx), \lle\bfsi_k^I\rle(\bfb_k,0,\bfx),
\nonumber \\
\!\!\!\!\!\!\!\!\!\!\!&&\!\!\! \lle{\bfep}_{ik}^I\rle(\bfb_k,0,\bfx),
\!\lle{\bfsi}_{ik}^I\rle(\bfb_k,0,\bfx), \bfb_k(\bfx), 0\},\nonumber \\
\label{4.7}
\!\!\!\!\!\!\!\!\!\!\!{\bfcD}^{II}&\!=\!&\{\bfcD^{II}_k\}_{k=1}^N, \ \ \ {\bfcD}^{II}_k=\{
\lle{\bfep}^{II}_k\rle({\bf 0},\theta_k,\bfx), \lle\bfsi_k^{II}\rle({\bf 0},\theta_k,\bfx),
\nonumber \\
\!\!\!\!\!\!\!\!\!\!\!&&\!\!\! \lle{\bfep}_{ik}^{II}\rle({\bf 0},\theta_k,\bfx),
\!\lle{\bfsi}_{ik}^{II}\rle({\bf 0},\theta_k,\bfx), {\bf 0},\theta_k(\bfx)\}.
\EEEQ

\vspace{2.mm} \noindent \hspace{10mm} \parbox{6.2cm}{
\centering \epsfig{figure=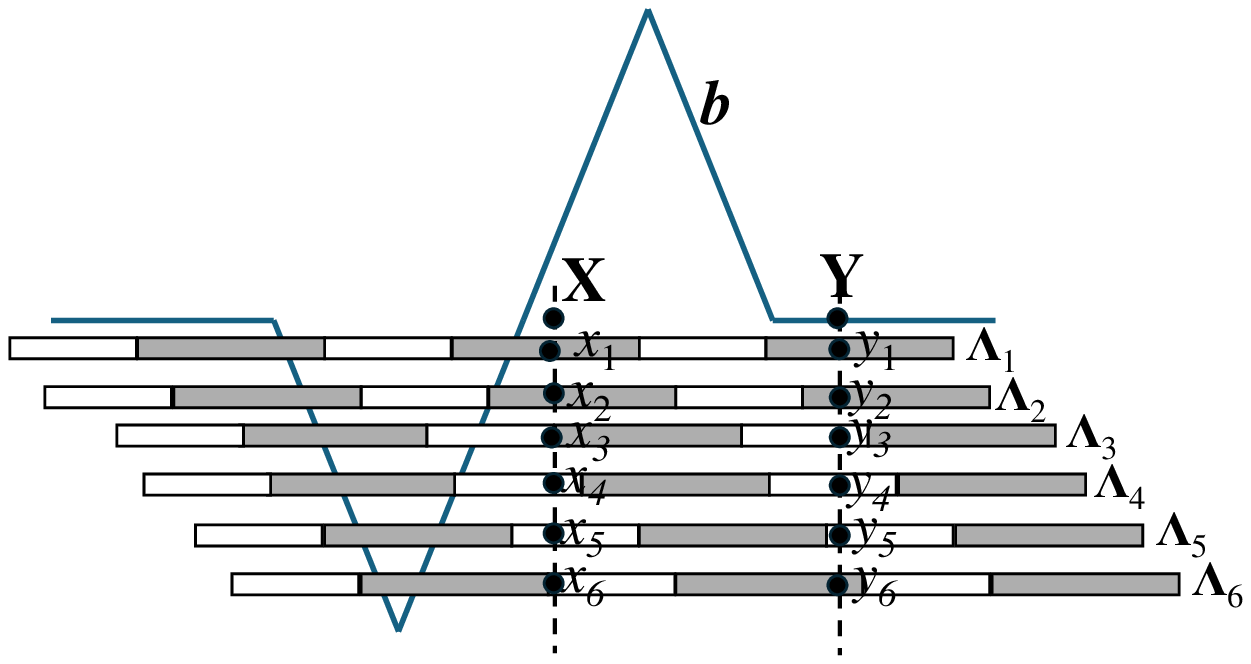, width=10.2cm}\\ \vspace{-22.mm}
\vspace{20.mm}
\vspace{-72.mm} \tenrm \baselineskip=8pt
{{\sc Fig. 4} Scheme of translated averaging}}
\vspace{2.mm}

\noindent 
Each effective dataset entry is defined as $\bfcD^{T}_k=({\bfcD_k^I,\bfcD^{II}_k})$, where $\bfcD_k^I$ and $\bfcD^{II}_k$ are computed for a particular realization of the loadings (\ref{3.11}) and (\ref{3.12}), respectively, during the offline stage using the averaging rules (\ref{4.5}) and (\ref{4.6}). It contains both macroscopic fields—average strain $\langle{\bfep}^T_k\rangle(\bfx)$ and stress $\langle{\bfsi}^T_k\rangle(\bfx)$—and microscopic inclusion fields $\langle{\bfep}^T_{ik}\rangle(\bfx):=\langle{\bfep}^T_k\rangle_i$, $\langle{\bfsi}^T_{ik}\rangle(\bfx):=\langle{\bfsi}^T_k\rangle_i$. Here, $\bfx \in \mathbb{R}^d$ denotes the macroscopic domain of the homogenized composite, so $\bfcD^T$ encapsulates only macroscopic-level information.
Although a sub-dataset $\bfcD^T_{\bfchi_ k}$—associated with a particular grid translation $\bfLa{\bf \chi}$—is obtained directly from DNS, the complete dataset $\bfcD^T_k$ is assembled through the translated averaging operations of (\ref{4.5}) and (\ref{4.6}). This averaging framework constitutes the core of computational analytical micromechanics (CAM), as formulated in \cite{{Buryachenko`2023}, {Buryachenko`2023a}}. Similarly, effective dataset $\bfcD^{\rm r}$ for random-structured composites, constructed by \cite{Buryachenko`2023} via the same CAM approach, is formally equivalent to the periodic effective dataset $\bfcD^T$ in (\ref{4.7}). Hence, the estimation of $\bfcD^T$ may also be referred to as the CAM method.
The resulting effective dataset $\bfcD^T$ provides a structured foundation for data-driven modeling, supporting efficient evaluation and interpolation of effective responses for new loadings (\ref{2.8}), (\ref{2.9}).

The construction of the effective dataset $\bfcD^T $ in (\ref{4.7}) may be carried out for either a finite body-force unit cell $\Omega_{00}^b$ or for the entire space $\Omega_{00}^b=\mathbb{R}^d$, provided the body force region ${\cal B}$ is compactly supported (\ref{2.5}), i.e., $B^b<\infty$.
By reformulating and extending the classical Definition 4.1 of \cite{Hill`1963}, one can establish a more general and adaptable definition that meets the needs of the present study, specifically for the case of a compact support loading $\bfb(\bfx)$ (\ref{2.8}) and
$\theta(\bfx)$ (\ref{2.9}) (see for details \cite{Buryachenko`2025c}):

\noindent {\bf Definition 4.2.} {\it An RVE is structurally fully representative of the entire composite material, ensuring that all apparent effective parameters $\bfcD^T_k$ ($k=1,\dots,N$) in (\ref{4.7}) are properly stabilized outside the region $\bfx \notin {\rm RVE}$. In other words, for an infinite periodic composite, both strains and stresses vanish in the exterior domain $\bfx \in \overline{\rm RVE}:=\Omega^b_{00}\setminus {\rm RVE}$.}

A central challenge in analyzing periodic CMs is selecting appropriate PBCs (\ref{2.15}). While these conditions are exact under homogeneous loading (\ref{2.13}) with zero CS loading (\ref{2.12}), they become inaccurate when nonzero body forces are present. In such cases, enforcing PBCs at the UC level is unnecessary, particularly with DNS. Instead, the RVE size is treated as a {\color{black} learned (or adjustable) parameter, chosen to satisfy a tolerance criterion }
\BBEQ
\label{4.8}
{\color{black} |\bfsi(\bfx)|,|\bfep(\bfx)|<{\rm tol}\ \ {\rm for}\ \ 
\forall\bfx\in\overline{\rm RVR}}
\EEEQ
{\color{black} that guarantees representativeness for homogenization. Moreover, since applied body forces $\{\bfb_k(\bfx)\}_{k=1}^N$ have compact support, the infinite-domain problem reduces to a finite RVE, eliminating finite-size and boundary (edge) effects. 
In fact, Eq. (\ref{4.8}) represents a learning parameter of vanishing condition implicitly introduced in \cite{Buryachenko`2023, Buryachenko`2023a, Buryachenko`2025c}, where the author presented a variety of numerical results for nonlocal effective deformations of a 1D peridynamic bar with a random structure, considering different scale ratios between material length scales (the inclusion size $a$ and the peridynamic horizon) and the field scale (the size of the compact support $B^b$). The numerical results explicitly demonstrate the elimination of both sample-size effects (owing to an infinite medium) and boundary-layer effects (due to the absence of boundaries).}

Definition 4.2 differs fundamentally from Definition 4.1. While Definition 4.1 addresses an idealized infinite domain with effective properties obtained via a limiting process, Definition 4.2 introduces a finite-size RVE, suitable for computation or experiment, based on different BCs (\ref{2.15}) and compact support loading (\ref{2.8}) and (\ref{2.9}). The focus is on stabilization of the effective dataset 
$\bfcD_k^p(\bfx)$ $(k=1,\ldots, N)$ 
(\ref{4.7}) in the exterior domain $\overline{\rm {RVE}}$ (\ref{4.8}), ensuring that effective parameters remain constant within tolerance in the annular region $B^{\rm RVE}\le|\bfx|\le B^{\rm RVE}+B^b/2$. When satisfied, the infinite medium can be represented by a finite RVE without edge effects. 
{\color{black} 
If $B^{\rm RVE}$ is too large for a particular $\bfb_k(\bfx)$ (e.g., RVE$\not\subset \Omega_{00}^b$), residual boundary artifacts and finite-size effects contaminate the corresponding particular effective dataset $\bfcD_k^p(\bfx)$ (\ref{4.7}), which must therefore be excluded from further consideration. This implies that the size $B^b$ (or the gradient magnitude $|\nabla \bfb(\bfx)|$)
(see Eq. (\ref{2.8})) should be reduced.}

With this adjustment, the domain of interest $\bfx \in w$ (see Fig. 5) is effectively restricted to the RVE, in line with Definition 4.2. The homogeneou remote loading (\ref{2.13}) of the original domain (Fig. 3) is replaced by compactly supported loading given in Eqs. (\ref{2.8}) and (\ref{2.9}), now illustrated in Fig. 5. In this setup, the localized force region $b(\bfx_i, B^b) \subset {\rm RVE}$ is explicitly shown, highlighting that external actions are confined within a subregion of the RVE. This confinement ensures stabilization of the effective parameters $\bfcD^T$ outside the RVE. The effective dataset $\bfcD^T$ is obtained from multiple realizations under different CS loadings (\ref{2.8}) and (\ref{2.9}) and distinct grid transition $\bfLa_{{\bf \chi}k}$. No restrictions are imposed on the geometry or size of either $b(\bfx_i, B^b)$ or the RVE; the spherical regions in Fig. 5 are chosen only for convenience. A numerical interpretation of Fig. 5 is provided in \cite{Buryachenko`2023, Buryachenko`2023a} for a randomly inhomogeneous bar under BFCS loading (\ref{2.8}).

Figure 4 shows the spatial hierarchy $\bfcB^b \subset {\rm RVE} \subset \Omega^b_{00}$ for composite materials (CMs) with two distinct microstructural configurations: periodic (Fig. 5a) and deterministic (Fig. 5b). In Fig. 5a, a representative periodic arrangement $X$ of inclusions is illustrated, specified by the set of centers $\bfLa_{\bf \chi}$. While the inclusion distribution is based on periodicity {\color{black} (material periodicity), the property of field periodicity} is not explicitly employed in the simulations. Instead, datasets $\bfcD^T_{{\bf \chi}_k}$ are constructed by applying multiple realizations ($k=1,\ldots, N$) of the loading (\ref{2.8}) and (\ref{2.9}) and computing the corresponding local fields within the BFUC $\Omega^b_{00}$ using appropriate numerical schemes (e.g., {\color{black}FEA} or FFT-based solvers described in \cite{Buryachenko`2025b}). The stress and strain fields are assumed to vanish outside the RVE (i.e., $\overline{\mathrm{RVE}}$), enabling the analysis of a finite collection of inclusions $v_i \subset \Omega^b_{00}$ without direct imposition of periodic boundary conditions (\ref{2.15}). Periodicity of $\bfLa_{\bf \chi}$ is invoked only during the translation-averaging procedure (see Eqs. (\ref{4.5}) and (\ref{4.6})).
Likewise, Fig. 5b depicts deterministic microstructures $X^j$ ($j = 1,2,\ldots$), which are neither periodic nor random, and may, for example, represent individual samples reconstructed from CT scans. As in the periodic setting, effective datasets {\color{black} $\bfcD^{\mathrm{d}j}_k$ are generated within $\Omega^b_{00}$} for each deterministic configuration $X^j$ under prescribed CS loading (\ref{2.8}), (\ref{2.9}). Collectively, these effective datasets form an ensemble-based representation, defined analogously to (\ref{4.7}) $\bfcD^{{\rm d}T}=(\bfcD^{{\rm d} I}, \bfcD^{{\rm d}II})$
\BBEQ
\!\!\!\!\!\!\!\!\!\!\!\bfcD^{{\rm d} I}&\!=\!&\{\bfcD^{{\rm d} I}_k\}_{k=1}^N, \ \ \ \bfcD^{{\rm d} I}_k=\{
\lle{\bfep}^I_k\rle(\bfb_k,0,\bfx), \lle\bfsi_k^I\rle(\bfb_k,0,\bfx),
\nonumber \\
\!\!\!\!\!\!\!\!\!\!\!&&\!\!\! \lle{\bfep}_{ik}^I\rle(\bfb_k,0,\bfx),
\!\lle{\bfsi}_{ik}^I\rle(\bfb_k,0,\bfx), \bfb_k(\bfx), 0\},\nonumber \\
\label{4.9}
\!\!\!\!\!\!\!\!\!\!\!\bfcD^{{\rm d}II}&\!=\!&\{\bfcD^{{\rm d}II}_k\}_{k=1}^N, \ \ \ \bfcD^{{\rm d}II}_k=\{
\lle{\bfep}^{II}_k\rle({\bf 0},\theta_k,\bfx), \lle\bfsi_k^{II}\rle({\bf 0},\theta_k,\bfx),
\nonumber \\
\!\!\!\!\!\!\!\!\!\!\!&&\!\!\! \lle{\bfep}_{ik}^{II}\rle({\bf 0},\theta_k,\bfx),
\!\lle{\bfsi}_{ik}^{II}\rle({\bf 0},\theta_k,\bfx), {\bf 0},\theta_k(\bfx)\}.
\EEEQ
Here, the statistical averaging $\lle \cdot \rle$ is carried out over the family of deterministic configurations $X^j$, thereby capturing the effective material response to each realization of the CS loading (\ref{2.8}), (\ref{2.9}) . No particular restrictions are imposed on the geometry of either the regions $\bfcB^b$ or the RVE (see Definition 4.2), nor on their relative dimensions. The spherical shapes depicted in Fig. 5 are chosen solely for illustrative clarity and do not represent any inherent geometric limitation.

\vspace{-1.mm} \noindent \hspace{-5mm} \parbox{16.2cm}{
\centering \epsfig{figure=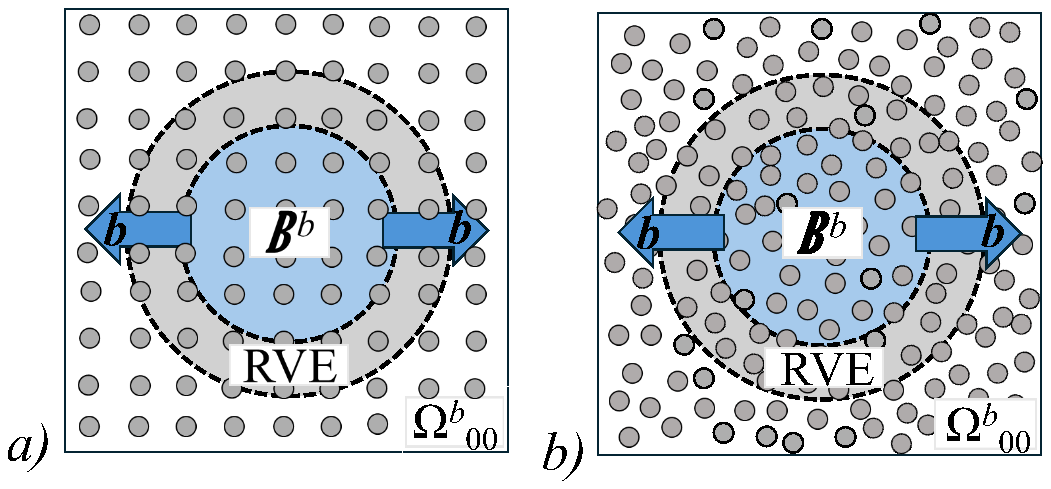, width=11.2cm}\\ \vspace{-22.mm}
\vspace{-72.mm} \tenrm \baselineskip=8pt
\centering{{\sc Fig. 5:} Schemes of $\bfcB^b\subset {\rm RVE}\subset\Omega^b_{00}$ for CM with periodic a) 
and deterministic b) structures}}
\vspace{1.mm}

A recent advance in computational micromechanics is presented by Silling {\it et al.} \cite{Silling`et`2024}, who propose a coarse-graining strategy for two-dimensional random composites. The microstructure, generated by Monte Carlo simulations in a square domain $w$ with $\sim$900 circular inclusions (Fig. 6a), assigns peridynamic behavior to each phase. While such systems are typically analyzed via classical RVE methods under Definition 4.1 (see Fig. 3) with periodic BC (\ref{2.15}), the authors instead apply a self-equilibrated body force $\bfb(\bfx)$ with compact support (\ref{2.8}) (similar to \cite{Buryachenko`2023, Buryachenko`2023a}). This redefines the RVE per Definition 4.2. The method yields negligible strains in the violaceous boundary layer $\overline{{\rm RVE}}=w\setminus{\rm RVE}$ (Fig. 6b); {\color{black} this is a best illustration of vanishing condition (\ref{4.8}). Though the term “RVE” is not explicitly used (this term was intrroduced later in \cite{Buryachenko`2025b})}, the multicolored region in Fig. 6b effectively represents it, while detailed fields $\lle\bfep\rle_i,\ \lle\bfsi\rle_i$ are deferred to future work.

\vspace{2.mm}
\hspace{-10mm} \parbox{12.8cm}{
\centering \epsfig{figure=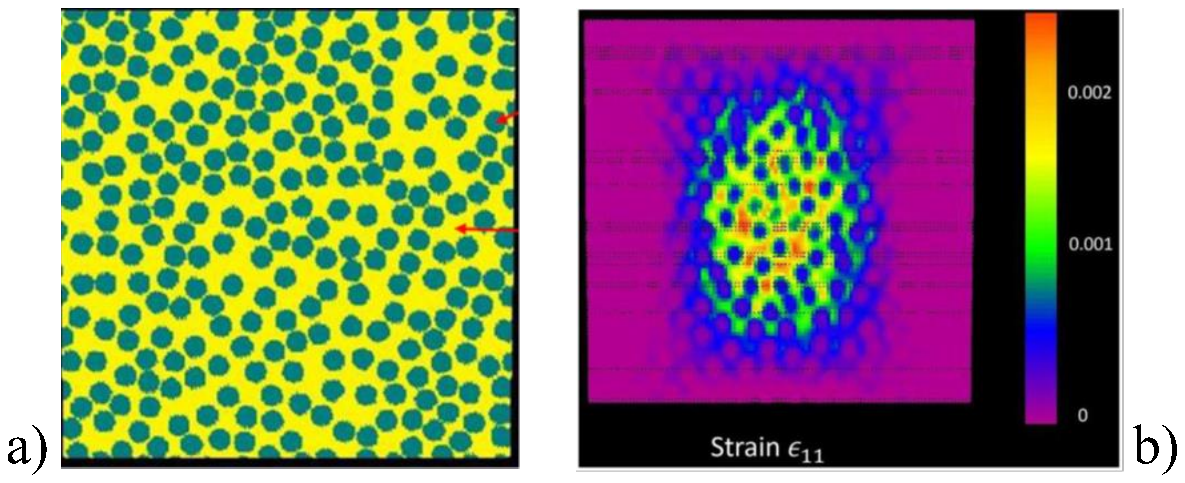, width=10.9cm}\\ \vspace{-22.mm}
\vspace{12.mm}
\vspace{-84.mm} \tenrm \baselineskip=8pt {\color{black}
{\hspace{5.mm}{\sc Fig. 6:} a) Simulated structure in $w$. b) DNS of strains $\bfep(\bfx)$ in $w$}}
}\vspace{1.mm}

Remarkably, the 1D results of \cite{{Buryachenko`2023},{Buryachenko`2023a}} (see the text after Eq. (\ref{3.6})) and the 2D case in Fig. 6b of \cite{Silling`et`2024}—though differing in dimension, microstructure, and numerical technique—are all naturally interpreted within the generalized RVE of Definition 4.2. This consistency across random \cite{Buryachenko`2023} and deterministic \cite{Silling`et`2024} models underscores the coherence, adaptability, and novelty of the proposed RVE concept—an uncommon but significant advance in micromechanics.

{\bf Comment 4.1.} 
It should be emphasized that the assumption of a matrix structure for the CMs is introduced only for the purpose of expressing the compact-support (CS) loading representations (\ref{3.4}) and (\ref{3.6}) in terms of the Green functions of the reference matrix (\ref{3.3}). This assumption serves as a convenient analytical device, rather than a fundamental limitation of the approach. In fact, when the formulation is implemented numerically—for example, within a finite element analysis (FEA) framework—the requirement of a matrix-type microstructure is not restrictive at all.
In particular, when evaluating the translation averages (\ref{4.5}) and (\ref{4.6}), one is not confined to matrix–inclusion morphologies but may instead employ any periodic microstructure (\ref{2.14}). This generality opens the door to a wide variety of material architectures, including skeletal frameworks, percolated networks, polycrystalline aggregates, and numerous other morphologies encountered in practice (see detailed discussion in \cite{Buryachenko`2007}). Consequently, the methodology retains both mathematical rigor and broad applicability, making it suitable for realistic modeling of diverse classes of composite systems.

{\color{black}
\noindent {\bf Comments 4.2.} 
It should be emphasized that a necessary condition for the emergence (and detection) of nonlocal effects is the inhomogeneity of the applied field and the comparability between the characteristic scale of this field and the intrinsic material scale (such as the inclusion size or the UC $|\Omega_{00}|$).
Inhomogeneity of the applied field may be introduced in two ways: a)
by prescribing a nonuniform remote loading (in contrast to the homogeneous loading of (\ref{2.8})), or b) by applying a body force b(x)as in $\bfb(\bfx)$, which need not coincide with the BFCS loading (\ref{2.8}).
Among these options, the BFCS loading (\ref{2.8}) is preferable for two reasons. Under the learning vanishing condition (\ref{4.8}), the problem for an infinite periodic medium (problem $\bfcR^{\rm per}$) reduces to the finite-size sample problem for an isolated domain (problem $\bfcR^{\rm iso}$; see Subsection 2.1), without requiring any additional boundary conditions (such as (\ref{2.13})).
Moreover, the problems $\bfcR^{\rm per}$ and f$\bfcR^{\rm iso}$ for domains $\Omega_{00}$ of arbitrary internal structure (see Comment 4.1) can be solved using any available numerical technique—for example, FEA, FFT, or the volume integral equation method (VIEM) (\ref{3.27})–(\ref{3.31}).
The solutions obtained on the inclusion grids $\bfLa_{\bf\chi}$ by any such method are then used within the translation-based averaging procedures (\ref{4.5}) and (\ref{4.6}), which are naturally well suited for parallel computing.
$\bfep(\bfx,\bfchi$ (\ref{4.5}), $\bfsi(\bfx,\bfchi$ (\ref{4.6}),
as well as the particular effective datasets $\bfcD^T_k$ and the total dataset $\bfcD^T$ (\ref{4.7}),
can be computed with arbitrarily high accuracy using any preferred numerical method.
Finally, the obtained effective (field) dataset
$\bfcD^T$ (\ref{4.7}) (or $\bfcD^{{\rm d}T}$ (\ref{4.9})) does not explicitly depend on the composite microstructure (e.g., on $c^{(1)}$ and $a/|\Omega_{00}|$), nor on the material properties of the individual phases or on the overall effective properties.
The computation of
$\bfcD^T$ (\ref{4.7}) (or $\bfcD^{{\rm d}T}$ (\ref{4.9})) 
essentially represents the completion of all underlying micromechanical problems.
The subsequent use of
$\bfcD^T$ (\ref{4.7}) (or $\bfcD^{{\rm d}T}$ (\ref{4.9}))
in Section 5 corresponds to its application within the macroscopic (effective) problem, although the effective concentration factors
$\lle{\bfep}_{ik}^I\rle(\bfb_k,0,\bfx)$ and 
$\!\lle{\bfsi}_{ik}^{I}\rle(\bfb_k,0,\bfx)$ (\ref{4.7})
will still appear in the formulation.
}

{\color{black} {\bf Comment 4.3.} Loosely speaking, the newly introduced concept of the effective dataset $\bfcD^T$ represents a dramatic leap beyond the classical notions of both the effective moduli $\bfL^*$ and the strain concentration factor $\bfA^*(\bfx)$. Let us compare the similarities and differences between the effective moduli $\bfL^*$ and the strain concentrator factor $\bfA^*(\bfz)$ (\ref{4.2}), on one hand, and the effective dataset $\bfcD^T$ (\ref{4.7}), on the other.
None of the quantities $\bfL^*$, $\bfA^*(\bfx)$, or $\bfcD^T$ depend explicitly on microstructural geometric descriptors (e.g., $c^{(1)}$) or phase properties. Moreover, $\bfcD^T$ (fielld parameter) does not depend on overall properties, unlike
$\bfL^*$ (material parameter). 
While $\bfL^*$, $\bfA^*(\bfx)$) are independent of the homogeneous boundary conditions (\ref{2.13}), the effective dataset $\bfcD^T$ (\ref{4.7}) depends explicitly on the CS loading (\ref{2.8}) 
and (\ref{2.9}). The key distinction, however, is that the effective moduli $\bfL^*$ form a single constant tensor, whereas the effective dataset $\bfcD^T=\{\bfcD^T_k\}_{k=1}^N$
can contain an arbitrary number of elements $\bfcD^T_k$ with potentially any nonlocal, or otherwise complex relationships among multiple effective parameters. As a result, the new concept of the effective dataset $\bfcD^T$ is far more informative than the constant tensor $\bfL^*$, while also eliminating both sample-size and boundary-layer effects.
It does not simply extend the idea of effective moduli $\bfL^*$ -- it fundamentally broadens their scope, capturing a far richer and more versatile description of material behavior than ever before. 
These properties make the effective dataset $\bfcD^T$ (\ref{4.7}) ideally suited for subsequent use in standard ML\&NN methodologies (see Section 5).
}

\section{Effective elastic moduli and surrogate operators }
\setcounter{equation}{0}
\renewcommand{\theequation}{5.\arabic{equation}}

Consider a periodic composite material with inclusion centers $\Lambda_{00}$ and subject to the periodic boundary conditions (PBC) (\ref{2.15}) on the unit cell $\Omega_{00}$. Applying the Gauss–Ostrogradsky theorem, we introduce the overall macrostress ${\bfsi} = \lle \bfsi \rle^{\Omega}$ and the overall macrostrain ${\bfep} = \lle \bfep \rle^{\Omega}$, both defined as averages over the unit cell $\Omega_{00}$
\BBEQ
\label{5.1}
\lle\bfsi\rle^{\Omega}\!\!\!&:=&\!\!\! |{\Omega_{00}}|^{-1}\int_{\Omega_{00}}\bfsi(\bfx)d\bfx=
|{\Omega_{00}}|^{-1}\int_{\Gamma^0}\bft(\bfs)\,\,\,^{^S}\!\!\!\!\!\! \otimes\bfs d\bfs,\\
\label{5.2}
\lle\bfep\rle^{\Omega}\!\!\!&:=&\!\!\! |{\Omega_{00}}|^{-1}\int_{\Omega_{00}}\bfep(\bfx)d\bfx=
|{\Omega_{00}}|^{-1}\int_{\Gamma^0}\bfu(\bfs)\,\,\,^{^S}\!\!\!\!\!\! \otimes\bfn(\bfs) d\bfs,
\EEEQ
in terms of the traction $\bft(\bfs) := \bfsi(\bfs)\bfn(\bfs)$ and displacement $\bfu(\bfs)$ on the geometric boundary of the unit cell $\bfs \in \Gamma^0$, with $\bfn(\bfs)$ denoting the outward unit normal to $\Gamma^0$. The effective tensor $\bfL^*$ and $\bfal^*$ is defined as the proportionality relation between the cell-averaged stress $\lle \bfsi \rle^{\Omega}$ and strain $\lle \bfep \rle^{\Omega}$
\BB
\label{5.3}
\lle\bfsi^I\rle^{\Omega}=\bfL^*\lle\bfep^I\rle^{\Omega}, \ \ \ \lle\bfsi^{II}\rle ^{\Omega}=\bfal^*.
\EE
The determination of these macroscopic quantities is achieved through the {{\it micro-to-macro}} transition. In particular, evaluations of the effective moduli via relation (\ref{5.3}) using MEA or FFT-based approaches represent a well-established line of research in micromechanics (see Introduction for references).

Our analysis now proceeds to the case of compact-support loading, governed by Eqs. (\ref{2.8}) and (\ref{2.9}).
A promising line in data-driven ML for composite materials (CMs) was opened by Silling \cite{Silling`2020} and advanced in \cite{You`et`2020, You`et`2024}. 
These studies developed surrogate nonlocal operators for infinite media using DNS of a finite 1D heterogeneous bar (periodic or random) under nonuniform loading: boundary wave excitation or internal oscillating body forces. The dataset is structured as
\BBEQ
\label{5.4}
\!\!\!\!\!\!\!\!\!\!\!{\bfcD}^{\rm DNS}=\{\bfcD^{\rm DNS}_k\}_{k=1}^N, \
\bfcD^{\rm DNS}_k=\{\bfu(\bfb_k,\bfx),\bfb_k(\bfx)\}.
\EEEQ
Each sample corresponds to a realization of $\bfb_k(\bfx)$, which need not satisfy the BFCS condition (\ref{2.8}). In contrast, the present study \cite{Buryachenko`2023, Buryachenko`2023a} extends the methodology to random microstructures (\ref{2.13}) by constructing structured effective datasets $\bfcD^{T}$ (\ref{4.1}) using BFCS (\ref{2.8}) and TCCS (\ref{2.9}) loadings. These loadings act as both excitation sources and controlled inputs for identifying nonlocal responses. The resulting datasets are compressed into reduced-order surrogates $\bfcD^{T}$, significantly lowering computational cost while retaining essential micromechanical features.

A related approach was introduced by Buryachenko \cite{Buryachenko`2022}, who formulated surrogate operators based on the concept of strongly nonlocal strain-type models, as pioneered by Eringen (see e.g., \cite{Eringen`2002}).
\BBEQ
\!\!\!\!\!\!\!\!\!\!\!\!\!\!\bfcL^{\epsilon T}_{\rm p}[\lle{\bfep}^T_k\rle](\bfx) &=& {\bf \Gamma}^T(\bfx), \nonumber\\
\label{5.5}
\!\!\!\!\!\!\!\!\! \!\!\!\!\!\bfcL^{\epsilon T}_{\rm p}[\lle{\bfep}^T_k\rle](\bfx) &=&
\!\!\int \!\!\bfK^{\epsilon T}_{\rm p}(|\bfx-\bfy|) \lle\bfep^{T}_k\rle(\bfy)~d\bfy,
\EEEQ
where the subscript ${\rm p}=b,\sigma,\epsilon_i,\sigma_i$ designates one of four distinct operator types, corresponding to either strain, stress, or their localized variants. The quantity ${\bf \Gamma}^T_k(\bfx) := ({\bf \Gamma}^I_k(\bfx), {\bf \Gamma}^{II}_k(\bfx))$ represents the averaged response fields—such as $\bfxi^{b(0)}$, $\langle \bfsi^I\rangle(\bfx)$, $\langle \bfep^I\rangle_i(\bfx)$, $\langle \bfsi^I\rangle_i(\bfx)$ for the mechanical case $\{\cdot\}^I$, Eq. (\ref{3.11})), or $\bfxi^{\theta}$, $\langle \bfsi^{II}\rangle(\bfx)$, $\langle \bfep^{II}\rangle_i(\bfx)$, $\langle \bfsi^{II}\rangle_i(\bfx)$ for the thermal case $\{\cdot\}^{II}$, Eq. (\ref{3.12})). These correspond to four types of learned surrogate operators:
$\bfcL^T_{\rm p} = ({ \bfcL^I_{\rm p}, \bfcL^{II}_{\rm p} })$.
The kernels $\bfK_{\rm p}^{\epsilon T*}$ are obtained by solving the 
minimization problem
\BBEQ
\label{5.6}
\!\!\!\!\!\!\!\!\!\!\!\!\!\!\!\!\!\!\!\!\!\bfK_{\rm p}^{\epsilon T*}={\rm arg}\!\min_{\!\!\!\!\!\!\!\!\!\! {{\bf K}_{\rm p}^{\epsilon T}}}
\!\sum_{k=1}^N\!|| \bfcL^{\epsilon T}_{\rm p}[\lle\bfep_k^T\rle](\bfx)- {\bf\Gamma}^T_k(\bfx)||^2_{l_2} 
+{\cal R}(\bfK_{\rm p}).
\EEEQ
where ${\cal R}(\bfK_{\rm p}^{\epsilon T})$
denotes a regularization functional--typically Tikhonov regularization—introduced to stabilize the ill-posed inverse problem. The kernel $\bfK_{\rm p}^{\epsilon T}$ is parameterized using a basis of Bernstein polynomials, and the optimization is carried out using the Adam optimizer \cite{Kingma`B`2014}. 
Earlier studies \cite{You`et`2024,You`et`2020,You`et`2021,You`et`2022} developed displacement-type surrogate operators (peridynamics) derived from datasets $\bfcD^{\rm DNS}$. By contrast, the compressed effective dataset ${\bfcD}^T$ (or ${\bfcD}^{{\rm d}T}$)—whether associated with periodic or deterministic microstructures—eliminate the need for full-field DNS. Instead, they rely on physically motivated loading protocols, systematic micromechanical averaging, and data-driven identification of nonlocal (strain-type) constitutive relations. This approach yields a flexible and robust framework for constructing surrogate operators applicable to a wide range of heterogeneous materials.

Under homogeneous loading conditions as in Eq. (\ref{2.10}) and constant temperature $\theta = {\rm const.}$, the effective properties are expressed through the surrogate operators 
\BBEQ
\label{5.7}
\lle\bfsi\rle&=&\bfL^{*}\lle\bfep\rle+\bfal^*,\ \ \bfL^*=\int(\bfK^{\epsilon I}_{\sigma})^{\top}(|\bfx-\bfy|)~d\bfy,\\
\label{5.8}
\bfal^*&=&-\int(\bfK^{\epsilon I}_{\sigma})^{\top}(|\bfx-\bfy|)~d\bfy\bfbe^{(0)}-\bar v_i^{-1}\int \int_{v_i}(\bfK^{\epsilon I}_{\lle\sigma\rle_i})^{\top}(|\bfx-\bfy|)\bfbe_1(\bfx)~d\bfx d\bfy.
\EEEQ

The surrogate operators introduced in \cite{Silling`2020}, \cite{You`et`2020, You`et`2024}, and (\ref{5.5}) are intrinsically static and primarily suited for modeling linear material behavior. To overcome these constraints, nonlocal neural operators have recently emerged as a powerful class of methods capable of learning generalized mappings between function spaces, thereby providing greater modeling flexibility and expressive capability \cite{Lanthaler`et`2024}, \cite{Li`et`2003}. In contrast, conventional neural network architectures such as fully connected neural networks (FCNNs) are fundamentally designed to approximate local nonlinear operators. Concretely, an $L$-layer FCNN $\Psi(\bfx): \bfR^{d_{\bf x}} \to \bfR^{d_{\bf u}}$ maps an input vector $\bfx$ to an output $\bfu$ via a sequence of successive transformations:
\BBEQ
\label{5.9}
\!\!\!\!\!\!\!\!\!\!\!\!\!\!\!\!\!\bfz^l(\bfx)=\bfcA({\bf w}^l \bfz^{l-1}(\bfx)+\bfb^l), \ \bfu(\bfx)={\bf w}^L \bfz^{L-1}(\bfx)+\bfb^L.
\EEEQ
Here, $\bfcA$ represents the activation function (e.g., ReLU or $\tanh$), while $({\bf w}^l, \bfb^l)_{l=1}^L$ denote the trainable parameters. Such networks remain inherently local, since the output at each location $\bfx$ depends only on the input features at that same point. In contrast, nonlocal neural operators explicitly incorporate spatial correlations by embedding domain-wide interactions into the architecture. This is typically achieved through kernel-based integral operators that aggregate information across the entire domain. A generic nonlocal layer can thus be expressed in the following form:
\BB
\label{5.10}
\bfz^l(\bfx)= \bfcA({\bf w}^l \bfz^{l-1}(\bfx)+\bfb^l+(\bfcK^l(\bfz^{l-1})(\bfx)).
\EE
In this formulation, $\bfcK^l$ denotes a learnable nonlocal kernel operator that aggregates information from neighboring points. Various neural operator architectures realize this principle in distinct ways, including Deep Operator Networks (DeepONets), PCA-Net, Graph Neural Operators, Fourier Neural Operators (FNOs), and Laplace Neural Operators (LNOs). The main differences among these approaches lie in the construction and application of the kernel $\bfcK^l$, which dictates the structure and extent of nonlocal interactions. For comprehensive comparisons and performance assessments across different application domains, see recent benchmark studies \cite{Kumara`Y`2023}, \cite{Lanthaler`et`2024}, \cite{Gosmani`et`2022}, \cite{HuZ`et`2024}.

The Peridynamic Neural Operator (PNO) \cite{Jafarzadeh`et`2024} 
and its heterogeneous variant HeteroPNO \cite{Jafarzadeh`et`2024b}
serve as a powerful surrogate operator $\bfcG$, approximating $\bfcG(\lle\bfu\rle)(\bfx) \approx -\bfb(\bfx)$, and is capable of capturing the behavior of highly nonlinear, anisotropic, and heterogeneous materials. In contrast to conventional models that depend on explicitly defined constitutive relations (e.g., Eq. (\ref{5.5})), the PNO enables data-driven modeling that offers both enhanced accuracy and computational efficiency.

{\color{black} Physics-Informed Neural Networks (PINNs) \cite{Raissi`et`2019}, \cite{Karniadakis`et`2021}, \cite{HuZ`et`2024}, \cite{Cuomo`et`2022}, \cite{Haghighata`et`2021}, \cite{Harandi`et`2024}, \cite{Kim`L`2024}, \cite{Ren`L`2024} advance neural network training by incorporating governing physical laws—such as Eq. (\ref{2.5})—into the loss functional through residual terms, thereby ensuring that the solutions remain consistent with the underlying physics. Integrating PINNs with neural operator architectures \cite{Faroughi`et`2024}, \cite{Gosmani`et`2022}, \cite{Wang`Y`2024} enables the unified treatment of nonlinear material behavior, microstructural heterogeneity, and nonlocal interactions within a hybrid data–physics framework. These approaches have demonstrated strong generalization performance, even in high-dimensional and multiscale contexts. Nonetheless, a major limitation persists: current models are generally restricted to finite computational domains, creating difficulties in directly addressing infinite or unbounded media, which are often required in micromechanics and homogenization studies.}

The micromechanical CAM model (outlined in Sections 4 and 5) can be naturally integrated into architectures such as, e.g., {\color{black} PNO \cite{Jafarzadeh`et`2024} and HeteroPNO \cite{Jafarzadeh`et`2024b} (comparative qualitative analysis of PINN and the approach proposed is presented in Section 6)} 
by replacing the conventional full-field dataset ${\bfcD}^{\rm DNS}$ (\ref{5.4}) with the compressed statistical dataset ${\bfcD^T}$
(\ref{4.7})
(or ${\bfcD}^{{\rm d}T}$ (\ref{4.9})). This substitution leads to the formulation of a CAM-based neural network (CAMNN). A key advantage of this approach is that it inherently eliminates boundary layer and size effects, which typically hinder the ability of standard neural operators to generalize across arbitrary domain geometries (see \cite{Jafarzadeh`et`2024}). Moreover, the CAMNN framework extends naturally to the entire space $\mathbb{R}^d$, removing the dependence on specific domain shapes or on residual losses associated with boundary conditions, as is common in several recent methods (e.g., \cite{{Eghbalpoor`S`2024},{Ning`et`2023},{Yu`Z`2024},{Yu`Z`2024b},{Zhou`Y`2024}}).
The essential benefit of employing ${\bfcD}^T$ rather than ${\bfcD}^{\rm DNS}$ lies in its construction for compact-support loading (\ref{2.8}) and (\ref{2.9}), with explicit incorporation of the RVE. This allows the definition of a nonlocal analog (\ref{5.5}) of the effective concentration tensor (\ref{4.2}$3$), thereby extending the CAMNN methodology to nonlinear processes such as fracture and plasticity. Importantly, preparing effective dataset at the RVE level removes boundary and size effects at their origin—artifacts that cannot be fully mitigated during training or post-processing. Without RVE-based preparation, these effects persist and compromise the fidelity of surrogate operators $\bfcG{\gamma}(\lle\bfu\rle)(\bfx)$ ($\gamma = b, \sigma, u_i, \sigma_i$, Eq. (\ref{5.5})). Consequently, the RVE framework is indispensable for achieving physically consistent and scale-invariant ML/NN predictions in computational micromechanics.

{\color{black}
\noindent {\bf Comments 5.1.} 
It should be stressed that the author does not claim to introduce a new method, nor to correct or modify any existing LM$\&$NN techniques. The author also does not assert that the new concept of the effective dataset
$\bfcD^T$ (\ref{4.7}) or $\bfcD^{{\rm d}T}$ (\ref{4.9}) 
can be readily incorporated into all LM$\&$NN frameworks.
What is stated, however, is that the existing LM$\&$NN approaches that rely on the use of
$\bfcD^{\rm DNS}$ (\ref{5.4})
(see, e.g., \cite{Silling`2020}, \cite{You`et`2020, You`et`2024}, \cite{Jafarzadeh`et`2024}, and HeteroPNO \cite{Jafarzadeh`et`2024b})
can be directly adapted by simply replacing
$\bfcD^{\rm DNS}\to \bfcD^{T}$ (or $\bfcD^{{\rm d}T}$). 
Naturally, any data-driven boundary-value problem must satisfy three essential components:
(1) material datasets that encode the constitutive relations at the material-point level,
(2) equilibrium, and (3) kinematic compatibility.
In the present framework, all of these ingredients are intrinsically embedded in the new effective dataset $\bfcD^{T}$ (see Comments 4.2, 4.3, end Conclusion).
Because these requirements are fully incorporated in advance, there is no need for explicit enforcement of the governing equations or boundary conditions in the subsequent conclusive data-driven stage $\bfcD^T\to$ML$\&$NN (\ref{5.5})–(\ref{5.10}).
All such issues have already been resolved in the preliminary construction of the effective dataset $\bfcD^{T}$ (see the Conclusion and Eq. (\ref{6.1})).
The elimination of sample-size and boundary-layer effects implies that, due to the learning-manageable condition (\ref{4.9}), the resulting nonlocal surrogate operators (see, e.g., Eq. (\ref{5.6})) are obtained—for the first time—for an infinite medium. These operators may then be applied to infinite media subjected to arbitrary loadings (not necessarily the CS loadings (\ref{2.8}) and (\ref{2.9})) and to any inhomogeneous remote boundary conditions, rather than only the homogeneous form (\ref{2.13}).
It is noteworthy that the mentioned property of the surrogate operators coincides with the corresponding property of the GIE (\ref{3.34}) (see the end of Subsection 3.4). In this sense, a surrogate operator may be regarded as a solution of the GIE. However, any modification of $\lle\bfep\rle(\bfx)$ requires solving the GIE (3.34) from scratch within a full micromechanical framework. In contrast, within the new $\bfcD^T\to$ML$\&$NN approach, all micromechanical computations are complited at the stage of estimating $\bfcD^T$, whereas subsequent analyses for arbitrary $\lle\bfep\rle(\bfx)$ become essentially instantaneous through the identified surrogate operators (see also Comment 4.2 and the central part of the Conclusion).

Naturally, the prediction of effective behavior in the vicinity of boundaries constitutes an additional and separate problem, which lies outside the scope of this paper. Interested readers may consult Subsection 13.5. in \cite{Buryachenko`2022}.

}

\section{ Conclusion}
\vspace{.0mm}
\setcounter{equation}{0}
\renewcommand{\theequation}{6.\arabic{equation}}

To clarify the essence of the proposed approach, we emphasize three aspects: the novelty of the problem formulation, the presented solution, and future directions—both theoretical and practical.
The CS loading (\ref{2.8}) and (\ref{2.9}) has not, to the author’s knowledge, been previously applied in micromechanics for either random or periodic structures, likely due to its underestimated significance. Its main advantage lies in enabling the estimation of unspecified surrogate nonlocal operators (see Section 5).
This is realized through several steps. First, CS loading defines a new RVE concept (Definition 4.2) via a general translation-based averaging procedure (\ref{4.5}), (\ref{4.6}). The required DNS for each grid $\bfLa_{\chi}$ can be obtained by any numerical method, particularly FFT-based solvers (see Section 3 and Comment 3.1). The construction of datasets $\bfcD^T$ (\ref{4.7}) and $\bfcD^{{\rm d}T}$ (\ref{4.9}) is critical, as they provide the input for ML\&NN frameworks (Section 5).
By itself, the Eq. (\ref{3.25}) ML-S equation (\ref{3.36}) under compact support loading has limited utility—except in special cases such as laser heating \cite{Buryachenko`2025}. However, when integrated with CS loading, the new AGIEs, the new RVE concept, refined effective dataset, and ML\&NN tools, it yields a practical predictive methodology:
\BB
\label{6.1}
{\rm BFCS}+{\rm TCCS}\ \to {\rm AGIE}\to \ {\rm DNS}\ \to \ {\rm RVE}\ \to\ \bfcD^T_{\chi k}\ \to\ \bfcD^T\ \to\ {\rm ML}\&{\rm NN}.
\EE
{\color{black} AGIE, the new RVE, and the effective dataset $\bfcD^T$ (and their incorporation into ML$\&$NN techniques) (\ref{6.1}) represent fundamentally new concepts. As such, the author is unable to point to direct prior publications in which these ideas were introduced by other authors. Namely, the combination of these concepts makes the approach (\ref{6.1}) the most fundamental step since the first foundations of micromechanics laid by
Poisson, Faraday, Mossotti, Clausius, Lorenz, and Maxwell (1824–1879) (see for details \cite{Buryachenko`2025}).}

A key advantage of CS loading (\ref{2.8}) and (\ref{2.8}) is its ability to generate effective datasets $\bfcD^T$ and $\bfcD^{{\rm d}T}$ while avoiding issues of sample size, boundary layers, and edge effects. In periodic CMs, the CAM framework solves problems (\ref{4.5}), (\ref{4.6}) on the RVE (possibly spanning several UCs) without enforcing PBCs (\ref{2.15}) at $\Omega_{00}$ interfaces. This makes the direct use of asymptotic homogenization \cite{{Bakhvalov`P`1984}, {Fish`2014}} and computational homogenization methods \cite{{Geers`et`2010}, {Kanout`et`2009}, {Kouznetsova`et`2001}, {Matous`et`2017}, {Raju`et`2021}, {Terada`K`2001}}—which inherently rely on PBCs (\ref{2.15})—questionable. Hence, the revised RVE concept (Definition 4.2), formulated under general CS loading (\ref{2.8}) and (\ref{2.8}), is central to the CAMNN framework and its ML\&NN extensions (Section 5) for both periodic and deterministic composites.

Significant progress has been made in constructing effective operators for random \cite{{Drugan`2000}, {Drugan`2003}, {Drugan`W`1996}} and periodic \cite{{Ameen`et`2018}, 
{Auriault`1983}, {Auriault`R`1993}, 
{Kouznetsova`et`2004a}, {Kouznetsova`et`2004b}, {Smyshlyaev`C`2000}} structures (see also \cite{Buryachenko`2022}). 
Nevertheless, these methods ({\color{black} based, in fact, on GIEs (\ref{3.33})-(\ref{3.35}) and their generalizations}) are confined to predefined operators—most commonly the classical fourth-order differential form. Departing from this framework, for instance by adopting a displacement-type strongly nonlocal model, would require reconstructing the entire micromechanical scheme from the ground up. ({\color{black} The proposed approach (\ref{6.1}), which is based on AGIEs (\ref{3.25}) in contrast to the aforementioned methods relying on GIEs, removes this restriction. The analysis is reduced to the construction of an effective dataset $\bfcD^T$ (or $\bfcD^{{\rm d}T}$), which is completely independent of any explicit description of the underlying microstructure, phase or overall material properties, and of the numerical solver employed (FEA, FFT, etc.).} 
All compressed effective datasets—$\bfcD^T$, $\bfcD^{{\rm d}T}$, or $\bfcD^{\rm r}$ {\color{black} (see for details \cite{Buryachenko`2025})} —are derived through averaging procedures applied to periodic (matrix, skeletal, percolated, or polycrystalline), deterministic, or random structures. Importantly, these effective datasets share a unified structure, regardless of the particular realization of the microstructure (e.g., $\bfLa_{\bf \chi}$). Once generated, the compressed effective dataset can be seamlessly approximated by ML$\&$NN techniques using either predefined or entirely “a priori” undefined surrogate models
{\color{black} eliminating dependence on sample size, boundary layer, or edge effects. 
Thus, micromechanical computations are required only once—for effective dataset construction—while all subsequent analyses are essentially instantaneous
(see also Comments 4.2).}

An emerging pathway to improved predictive capability is the use of ML\&NN methods for constructing surrogate operators. However, these approaches often overlook essential micromechanical principles such as scale effects, boundary layers, and the RVE concept. The proposed CAMNN framework addresses this gap by producing compressed effective datasets for both periodic and deterministic microstructures.
Its novelty lies in replacing the DNS dataset $\bfcD^{\rm DNS}$ (\ref{5.4}) with $\bfcD^T$ (\ref{4.7}) or $\bfcD^{{\rm d}T}$ (\ref{4.9}), while preserving the Block Optimization step (\ref{5.6}). At the core is a new RVE definition (Definition 4.2), based not on phase constitutive laws or operator form but on field concentration factors, making it broadly applicable and model-independent. These effective datasets embed the generalized RVE concept, enabling seamless integration into ML\&NN frameworks for predicting nonlocal surrogate operators. {\color{black} Owing to learning vanishing conditions (\ref{4.8})
eliminating size effects, boundary layers, and edge effects}, CAMNN ensures both reliability and generalizability.
Furthermore, CAMNN accommodates linear and nonlinear behaviors, coupled or uncoupled elasticity, and both weakly nonlocal (e.g., gradient-type) and strongly nonlocal (e.g., strain-based, displacement-based, peridynamic) constitutive models \cite{Maugin`2017}. In this way, AGIE-CAM moves beyond conventional micromechanics, providing a unified, extensible framework for next-generation material modeling (see foe details {\cite{Buryachenko`2025}).

Thus, a condensed representation of the scheme (\ref{6.1}) underlying the proposed approach is shown in Fig. 7. Block 1 contains the input description of the BFCS (\ref{2.8}) and \ (\ref{2.9}). The central Block 2 is responsible for the evaluation of the effective dataset $\bfcD^T$ (or $\bfcD^{{\rm d}T}$).
It is important to note that these effective datasets $\bfcD^T$ carry no information about the microstructure of the composite material—such as, e.g., inclusion concentration $c^{(1)}$ and geometric ratios like $a/|\Omega_{00}|$, or the local/nonlocal constitutive properties of individual phases. Likewise, they remain independent of the computational method employed (e.g., FEA, VIEM, or FFT). At the same time, all size effects, boundary-layer influences, and edge artifacts are completely eliminated {\color{black}(see for details Comments 4.2, 4.3, 5.1).}

Finally, the effective datasets $\bfcD^T$ (or $\bfcD^{{\rm d}T}$) are transferred to Block 3, the ML$\&$NN module. {\color{black} In standard practice, the PINN (and its variants) 
performs training by minimizing residuals of PDEs, boundary conditions, and initial conditions, with a known material model. This makes PINNs (enforcing physical consistency) model-driven rather than data-driven, see \cite{Raissi`et`2019, Karniadakis`et`2021, Hu`et`2024}).
In the present approach (\ref{6.1}), however, such explicit enforcement of governing- and boundary-condition losses becomes unnecessary in the conclusive step
$\bfcD^T\to $ML$\&$NN (\ref{6.1})
because these constraints are already {\it explicitely} embedded in the effective dataset $\bfcD^T$ (or $\bfcD^{{\rm d}T}$) itself (see Eq. (\ref{6.1}) and Comments 4.2 and 5.1).
Thus, both PINNs and the present total version of the approach (\ref{6.1}) are Physics-Informed, but the nature of the physics enforcement is conceptually different: in the proposed method (\ref{6.1}), the physics is satisfied before the optimization problem (e.g., Eq. (\ref{5.6})).
In this way, the Physics-Informed features embedded in $\bfcD^T$ are evaluated at the highest achievable fidelity, with any desired level of numerical accuracy (see Comment 4.2).
It would be of interest to compare the computational efficiency of these two fundamentally different strategies. Nevertheless, the clear priority in eliminating boundary-layer effects remains with the present approach (\ref{6.1}) (see Comments 4.2 and 5.1).
Becauseof this, incorporation of $\bfcD^T$ into the PNO and HeteroPNO (see the end of Section 5) by straightforward replacing 
$\bfcD^{\rm DNS} \to\bfcD^T$ is presented more prefarable than incorporation of $\bfcD^T$ into PINN.
}

Moreover, the internal details of Block 2 are irrelevant to Block 3, and vice versa. In this sense, Blocks 2 and 3 may be regarded as “black boxes” with respect to each other (shown in Fig. 7 in black). Thus, for the first time, the methodology is presented in the form of a formalized modular scheme (Fig. 7), where experts developing one block need not possess expertise in the underlying principles of the other. Furthermore, in joint software development, each team (responsible for either Block 2 or Block 3) can modify or refine its own module independently at any stage, without requiring detailed consultation with the partner team. Only high-level coordination—to ensure smooth interfacing and data exchange between Blocks 2 and 3—is necessary for effective collaboration.

\vspace{0.mm}
\hspace{2mm} \parbox{12.8cm}{
\centering \epsfig{figure=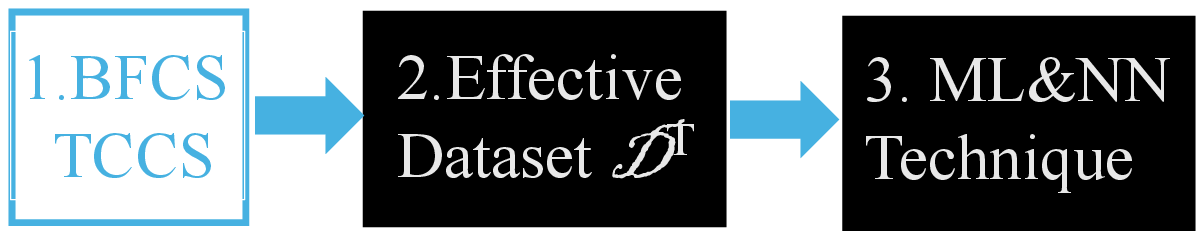, width=9.cm}\\ \vspace{-22.mm}
\vspace{-7.mm}
\vspace{-67.mm} \tenrm \baselineskip=8pt {\color{black}
{\hspace{5.mm}{\sc Fig. 7:} Scheme of proposed approach}}
}\vspace{1.mm}

\smallskip
\noindent{\bf Acknowledgments:}

{\color{black} The author acknowledges Dr. Stewart A. Silling for the helpful
comments, fruitful personal discussions, and encouragement suggestions.
The author acknowledges the reviewers for the
encouraging comments that initiated a significant correction of the manuscript.}
Permission to reproduce Figs. 6a and 6b from Silling {\it et al.} \cite{Silling`et`2024} was granted by Springer Nature (License Number 6044880317237).



\end{document}